\documentclass[conference]{IEEEtran}

\renewcommand\IEEEkeywordsname{Index Terms}
\usepackage{graphicx}
\usepackage{subcaption}
\usepackage{tikz,xparse}
\usetikzlibrary{dsp,chains}
\usetikzlibrary{matrix}
\usepackage{mathptmx}
\usepackage{verbatim}
\usepackage{calc}
\usepackage{ifthen}
\usepackage{xifthen}
\usepackage{cancel}
\usepackage{bm}
\usepackage{verbatim}
\usepackage{multirow}
\usepackage{cite}

\usepackage[nolist]{acronym} 
\usepackage{pgfplots}
\usetikzlibrary{arrows,shapes,graphs,graphs.standard,quotes}
\usetikzlibrary{matrix}
\pgfplotsset{compat=newest}
\usepackage[bookmarks=false]{hyperref}
\usepackage{units}
\usepackage{amsmath, amsbsy, amssymb, latexsym }
\hypersetup{bookmarksdepth=-2}
\usepackage{comment}
\usepackage[utf8]{inputenc}
\usepackage{xcolor}
\usepackage{enumitem}
\usepackage{algorithm}
\usepackage{algpseudocode}
\usetikzlibrary{spy}

\renewcommand{\Comment}[2][.75\linewidth]{
  \leavevmode\hfill\makebox[#1][l]{$\triangleright$~#2}}
\tikzset{>=latex}

\algnewcommand\algorithmicforeach{\textbf{for each}}
\algdef{S}[FOR]{ForEach}[1]{\algorithmicforeach\ #1\ \algorithmicdo}

\algnewcommand\algorithmicswitch{\textbf{switch}}
\algnewcommand\algorithmiccase{\textbf{case}}
\algnewcommand\algorithmicassert{\texttt{assert}}
\algnewcommand\Assert[1]{\State \algorithmicassert(#1)}

\algdef{SE}[SWITCH]{Switch}{EndSwitch}[1]{\algorithmicswitch\ #1\ \algorithmicdo}{\algorithmicend\ \algorithmicswitch}
\algdef{SE}[CASE]{Case}{EndCase}[1]{\algorithmiccase\ #1}{\algorithmicend\ \algorithmiccase}
\algtext*{EndCase}

\definecolor{mittelblau}{RGB}{0, 126, 198}
\definecolor{violettblau}{cmyk}{0.9, 0.6, 0, 0}
\definecolor{rot}{RGB}{238, 28 35}
\definecolor{apfelgruen}{RGB}{140, 198, 62}
\definecolor{gelb}{RGB}{255, 229, 0}
\definecolor{orange}{RGB}{244, 111, 33}
\definecolor{pink}{RGB}{237, 0, 140}
\definecolor{lila}{RGB}{128, 10, 145}
\definecolor{hellgrau}{RGB}{224, 224, 224}
\definecolor{mittelgrau}{RGB}{128, 128, 128}
\definecolor{dunkelgrau}{RGB}{80,80,80}
\definecolor{anthrazit}{RGB}{19, 31, 31}
\definecolor{darkgreen}{RGB}{34,139,34}
\colorlet{Mycolor1}{green!10!orange!90!}
\tikzset{
       vnd/.style={
        shape=circle,
        fill=black,
        draw,
        inner sep=0pt,
        minimum size=0.2cm},
        cnd/.style={
        shape=rectangle,
        fill=white,
        draw,
        minimum width=0.05mm,
        minimum height = 0.05mm}, 
         vndR/.style={
        shape=circle,
        fill=red,
        draw,
        inner sep=0pt,
        minimum size=0.2cm},
        cndR/.style={
        shape=rectangle,
        fill=white,
        draw=red,
        minimum width=0.05mm,
        minimum height = 0.05mm}
}

\IEEEoverridecommandlockouts

\def \reviewmode{\if01}

\renewcommand{\vec}[1]{\mathbf{#1}}

\newcommand{\rv}{\vec{r}}

\newcommand{\uv}{\vec{u}}

\newcommand{\xv}{\vec{x}}
\newcommand{\yv}{\vec{y}}

\newcommand{\Gm}{\vec{G}}

\begin{document}
\begin{NoHyper}
\title{Decoder-tailored Polar Code Design Using the Genetic Algorithm}

\iftrue \reviewmode
	\author{\IEEEauthorblockN{Author 1, Author 2, Author 3 and Author 4}
	\IEEEauthorblockA{}}
\else
	\author{\IEEEauthorblockN{Ahmed Elkelesh, Moustafa Ebada, Sebastian Cammerer and Stephan ten Brink}\thanks{Manuscript submitted September 20, 2018; revised January 28, 2019; date of current version January 28, 2019.
			
			The authors are with the Institute of Telecommunications, Pfaffenwaldring 47, University of Stuttgart, 70569 Stuttgart, Germany (e-mail: {elkelesh,ebada,cammerer,tenbrink}@inue.uni-stuttgart.de).

This work has been supported by DFG, Germany, under grant BR 3205/5-1.

Parts of this work have been accepted in the \emph{International ITG Conference on Systems, Communications and Coding (SCC), Feb. 2019} \cite{GenAlgSCC19}.} 
	\IEEEauthorblockA{
}
}

\fi

\maketitle

\begin{acronym}
 \acro{ECC}{error-correcting code}
 \acro{HDD}{hard decision decoding}
 \acro{SDD}{soft decision decoding}
 \acro{ML}{maximum likelihood}
 \acro{GPU}{graphical processing unit}
 \acro{BP}{belief propagation}
 \acro{BPL}{belief propagation list}
 \acro{LDPC}{low-density parity-check}
  \acro{HDPC}{high density parity check}
 \acro{BER}{bit error rate}
 \acro{SNR}{signal-to-noise-ratio}
 \acro{BPSK}{binary phase shift keying}
 \acro{AWGN}{additive white Gaussian noise}
 \acro{MSE}{mean squared error}
 \acro{LLR}{Log-likelihood ratio}
 \acro{MAP}{maximum a posteriori}
 \acro{NE}{normalized error}
 \acro{BLER}{block error rate}
 \acro{PE}{processing elements}
 \acro{SCL}{successive cancellation list}
 \acro{SC}{successive cancellation}
 \acro{BI-DMC}{Binary Input Discrete Memoryless Channel}
 \acro{CRC}{Cyclic Redundancy Check}
 \acro{BEC}{Binary Erasure Channel}
 \acro{BSC}{Binary Symmetric Channel}
 \acro{BCH}{Bose-Chaudhuri-Hocquenghem}
 \acro{RM}{Reed--Muller}
 \acro{RS}{Reed-Solomon}
  \acro{SISO}{soft-in/soft-out}
\acro{PSCL}{partitioned successive cancellation list}
  \acro{3GPP}{3rd Generation Partnership Project }
  \acro{eMBB}{enhanced Mobile Broadband}
      \acro{CN}{check nodes}
      \acro{PC}{parity-check}
      \acro{GenAlg}{Genetic Algorithm}
\acro{AI}{Artificial Intelligence}
\acro{MC}{Monte Carlo}
\acro{CSI}{Channel State Information}
\acro{PSCL}{partitioned successive cancellation list}
\end{acronym}

\begin{abstract}

We propose a new framework for constructing polar codes (i.e., selecting the frozen bit positions) for arbitrary channels, and tailored to a given decoding algorithm, rather than based on the (not necessarily optimal) assumption of \ac{SC} decoding.
The proposed framework is based on the \ac{GenAlg}, where populations (i.e., collections) of information sets evolve successively via evolutionary transformations based on their individual error-rate performance. These populations converge towards an information set that fits both the decoding behavior and the defined channel.
Using our proposed algorithm over the \ac{AWGN} channel, we construct a polar code of length $2048$ with code rate $0.5$, without the CRC-aid, tailored to \emph{plain} \ac{SCL} decoding, achieving the same error-rate performance as the CRC-aided SCL decoding, and leading to a coding gain of $\unit[1]{dB}$ at BER of $10^{-6}$. Further, a \ac{BP}-tailored construction approaches the \ac{SCL} error-rate performance without any modifications in the decoding algorithm itself. The performance gains can be attributed to the significant reduction in the total number of low-weight codewords. To demonstrate the flexibility, coding gains for the Rayleigh channel are shown under \ac{SCL} and \ac{BP} decoding.
Besides improvements in error-rate performance, we show that, when required, the GenAlg can be also set up to reduce the decoding complexity, e.g., the \ac{SCL} list size or the number of \ac{BP} iterations can be reduced, while maintaining the same error-rate performance.

\end{abstract}

\begin{IEEEkeywords}
	Polar Codes, Channel Polarization, Polar Code Construction, Reed--Muller Codes, Genetic Algorithm, Evolutionary Algorithms, Artificial Intelligence.
\end{IEEEkeywords}

\acresetall
\vspace{-0.2cm}
\section{Introduction}

Polar codes \cite{ArikanMain} are the first family of codes proven to be capacity achieving for any \ac{BI-DMC} and with an explicit construction method under low complexity \ac{SC} decoding. However, for finite lengths, both the \ac{SC} decoder and the polar code itself (i.e., its \ac{ML} performance) are shown to be sub-optimal when compared to other state-of-the-art coding schemes such as \ac{LDPC} codes \cite{polar_comparisons}. Later, Tal and Vardy introduced a \ac{SCL} decoder \cite{talvardyList} enabling a decoding performance close to the \ac{ML} bound for sufficiently large list sizes. A complexity reduced version was proposed in \cite{miloslavskaya2014sequential} with a slight negligible performance degradation. The concatenation with an additional high-rate \ac{CRC} code \cite{talvardyList} or \ac{PC} code \cite{PCC} further improves the code performance itself, as it increases the minimum distance and, thus, improves the weight spectrum of the code. This simple concatenation renders polar codes into a powerful coding scheme. For short length codes \cite{liva2016}, polar codes were recently selected by the 3GPP group as the channel codes for the upcoming \emph{5th generation mobile communication standard (5G)} uplink/downlink control channel \cite{polar5G2018}. 

On the other hand, some drawbacks can be seen in their high decoding complexity and large latency for long block lengths under \ac{SCL} decoding due to their inherent sequential decoding manner (see \cite{HW} and references therein), which currently limits the usage of polar codes in several practical applications. 

Although several other decoding schemes such as the iterative \ac{BP} decoding \cite{ArikanBP} and the iterative \ac{BPL} decoder \cite{elkelesh2018belief} exist, they are currently not competitive with \ac{CRC}-aided \ac{SCL} decoding in terms of error-rate performance \cite{polarCodes_concepts}. However, they promise more efficient decoder architectures in terms of decoding latency and parallelism. The \ac{PSCL} decoder \cite{PSCL_Journal} was introduced achieving almost the same performance of the \ac{SCL} decoder with a significantly lower memory requirement. Therefore, a good polar code design tailored to an explicit decoder may change this situation as it promises either an improved error-rate performance or savings in terms of decoding complexity for a specific type of decoder. 
In this work, we show that both can be achieved by considering the decoding algorithm throughout the code construction phase instead of constructing the code based on the typical assumption of \ac{SC} decoding. 
An intuitive example why a design for \ac{SC} is sub-optimal for other decoding schemes can be given by considering the girth of the graph under \ac{BP} decoding, which strongly depends on the frozen positions of the code. Thus, freezing additional nodes can even result in a degraded decoding performance under \ac{BP} decoding although the \ac{ML} performance indeed gets better (see Fig.~9 in \cite{ISWCS_Error_Floor}).

In a strict sense, one may argue that polar codes are inherently connected with \ac{SC} decoding and only a design based on the concept of channel polarization results in a \emph{true} ``polar'' code (cf. \ac{RM} codes). However, from a more practical point of view, we seek to find the most efficient coding scheme for finite length constraints and, with slight abuse of notation, we regard the proposed codes as polar codes. Thus, a design method that considers the decoder and improves the overall performance is an important step for future polar code applications.

Except for the \ac{BEC}, existing polar code constructions rely either on analytic approximations/bounds such as the Bhattacharyya parameter \cite{ArikanMain}, density evolution \cite{constructDE} and Gaussian approximation \cite{constructGaussian,GA} or heuristics \cite{PW,TUM_SCL_Construct, BetaIngmard}, including Monte-Carlo-based simulations for specific decoding algorithms \cite{MC_BP, MC_BP_2,BP_LLR_Siegel}. Additionally, several concatenation schemes of polar codes and other coding techniques have been proposed \cite{HybridTse, subCodes, BP_felxible, BP_Siegel_Concatenating, BP_sEXIT} to improve the finite length performance under different decoders. However, an explicit design tailored to \ac{SCL} or \ac{BP} decoding turns out to be cumbersome due to the many dependencies in the decoding graph and the high dimensionality of the optimization problem. 

Polar code construction (or design), throughout this paper, refers to selecting an appropriate frozen bit position vector. In this work, we propose a new framework for polar code construction matched to a specific decoding algorithm embedded in the well-understood \ac{GenAlg} context. As a result, the optimization algorithm works on a specific error-rate simulation setup, i.e., it inherently takes into account the actual decoder and channel. This renders the \ac{GenAlg}-based polar code optimization into a solid and powerful design method. Furthermore, to the best of our knowledge, the resulting polar codes in this work outperform any known design method for \emph{plain} polar codes under \ac{SCL} decoding \emph{without} the aid of an additional \ac{CRC} (i.e., CRC-aided SCL performance could be achieved \emph{without} the aid of a \ac{CRC}). Additionally, the \ac{BP} decoder of the \emph{proposed} code achieves (and slightly outperforms) the \ac{SCL} decoding performance of the \emph{conventional} code without any required decoder modifications. We noticed that the performance gains can be attributed to the significant reduction in the total number of low-weight codewords. 
Furthermore, the decoding complexity, latency and memory requirements can be reduced for a fixed target error-rate after carefully constructing the polar code using \ac{GenAlg}. We make the source code public and also provide the best polar code designs from this work online\footnote{\url{Link-will-be-Available-After-Review}}. Other optimization algorithms (e.g., differential evolution) can be used to solve the polar code design problem. However, due to the hard-decision nature of the problem (i.e., a bit-channel can be either frozen or non-frozen) GenAlg seems to be more suitable.

The paper is organized as follows. In Sec. \ref{sec:polarcodes}, we briefly review the fundamental concepts of polar codes. In Sec. \ref{sec:con},
the problem of polar code construction is presented along with the most relevant work. Sec. \ref{sec:genAlg} introduces the \ac{GenAlg} and discusses its preliminary concepts. The \ac{GenAlg}-based polar code construction is then presented in Sec. \ref{sec:genAlgcon}.
Results are presented in Sec. \ref{sec:resAWGN} and Sec. \ref{sec:resRayleigh} for \ac{AWGN} and Rayleigh fading channels, respectively. In Sec. \ref{sec:complexity}, we show that the improved code construction method can lead to significant reduction in decoding complexity, latency and memory requirements. Sec. \ref{sec:conc} depicts some conclusions. 

\section{Polar codes} \label{sec:polarcodes}

Polar codes \cite{ArikanMain} are based on the concept of channel polarization, in which $N=2^n$ identical copies of a channel $W$ are combined and $N$ synthesized bit-channels are generated. These synthesized bit-channels show a polarization behavior, in the sense that some bit-channels are purely noiseless and the rest are completely noisy.
A recursive channel combination provides the polarization matrix 
\begin{align*} 
\mathbf{G}_N = \mathbf{B}_N \cdot \mathbf{F}^{\otimes n}, \qquad \mathbf{F} = \left[ \begin{array}{ll} 1 & 0 \\ 1 & 1 \end{array}\right] 
\end{align*}
where $\mathbf{B}_N$ is a bit-reversal permutation matrix and $\mathbf{F}^{\otimes n}$ denotes the $n$-th Kronecker power of $\mathbf{F}$. Extensions for kernels other than the $2\times2$ kernel exist, but are not considered in this work. However, the results from this work can be applied straightforwardly.
The polar codewords $\xv$ are given by $\xv = \uv \cdot \Gm_N$, where $\mathbf{u}$ contains $k$ information bits and $N-k$ frozen bits, w.l.o.g. we set the frozen positions to ``0''. 
The information set $\mathbb{A}$ contains the $k$ most reliable positions of $\uv$ in which the $k$ information bits are transmitted and $\bar{\mathbb{A}}$ denotes the frozen positions (i.e., the complementary set to $\mathbb{A}$). The \emph{conventional} generator matrix, denoted by $\Gm$, is constructed as the rows $\left\{\rv_i\right\}$ of $\Gm_N$ with $i\in\mathbb{A}$.
The task of the polar code construction, in its original form, is to find the information set $\mathbb{A}$ which maximizes the code performance (under \ac{SC} decoding) for a specific channel condition. More details on the problem of polar code construction is provided in Sec. \ref{sec:con}.

In this work, a polar code with codeword length $N$ and $k$ information bits is denoted by $\mathcal{P} \left(N,k\right)$, i.e., the information set has the cardinality $|\mathbb{A}|=k$ and the code rate $R_c = k/N$.

Systematic polar encoding \cite{arikan2011systematic} can be applied which enhances the \ac{BER} performance with the same \ac{BLER} performance, when compared to non-systematic polar codes. 
Throughout this work, we use non-systematic polar encoding. However, it is straightforward to use the \ac{GenAlg} to construct systematic polar codes.

The basic polar decoding algorithms are:
\begin{itemize}
\item \textbf{\ac{SC} decoding} \cite{ArikanMain}; the first proposed decoder for polar codes, where all information bits $\hat{u}_i$ are sequentially hard-decided based on the previously estimated bits, $\left\{\hat{u}_1,\dots,\hat{u}_{i-1}\right\}$ and the channel information $\yv$, where $i\in\left\{1,\dots,N\right\}$. Obviously, it suffers from unavoidable error propagation (i.e., decision error at decided bit $\hat{u}_i$ cannot be corrected later and will eventually affect all next bit decisions).
\item \textbf{\ac{SCL} decoding} \cite{talvardyList}; denoted by \ac{SCL} $\left(L\right)$, utilizes a list of $L$ most likely candidate paths during \ac{SC} decoding; at every decision the decoder branches into two paths ($\hat{u}_i=0$ and $\hat{u}_i=1$) instead of the hard decision in the \ac{SC} decoder. To limit the exponential growth of complexity, only the $L$ most reliable paths are kept sorted in the list according to a specific path metric. 
\item \textbf{CRC-aided SCL decoding} \cite{talvardyList}; denoted by \ac{SCL}+\ac{CRC}-$r$ $\left(L\right)$, where an additional high-rate \ac{CRC} of $r$ bits is concatenated to the polar code, to help in selecting the final codeword from the $L$ surviving candidates, yielding significant performance gains in competing with the state-of-the-art error correcting codes.
\item \textbf{\ac{BP} decoding} \cite{ArikanBP}; denoted by \ac{BP} $\left(N_{it,max}\right)$ is an iterative message passing decoder based on the \emph{encoding} graph of the polar code. \ac{LLR} messages are iteratively passed along the encoding graph until reaching a maximum number of iterations $\left(N_{it,max}\right)$ or meeting an early stopping condition. It can be inherently parallelized and allows soft-in/soft-out decoding. However, the error-rate performance of polar codes under \ac{BP} decoding is typically not competitive with state-of-the-art \ac{SCL} decoding. Throughout this work, we use the \emph{$\mathbf{G}$-matrix-based} early stopping condition (``re-encoding''), where decoding terminates when $\mathbf{\hat{x}} = \mathbf{\hat{u}}\cdot\mathbf{G}_N$ \cite{earlyStop}.
\end{itemize}

\section{Polar Code Construction} \label{sec:con}

The polar code construction phase is about deciding the most reliable $k$ bit positions that are set as the information bit positions, while the remaining $N-k$ bit positions are set as frozen bit positions. Thus, ranking the bit-channels according to their reliabilities is of major significance in the polar code construction phase. 
The information set $\mathbb{A}$ is the outcome from the polar code construction phase specifying the indices of the information bit positions. A corresponding logical $\mathbf{A}$-vector can be used such that $\mathbf{A} = \left[ a_1,a_2,\dots,a_N \right]$, where $a_i \in \left \{ 0,1 \right \}$ and $1 \le i \le N$. Bit position $i$ is frozen if $a_i = 0$, while bit position $j$ is non-frozen (i.e., can be used for information transmission) if $a_j = 1$. For instance consider the $\mathcal{P} \left(8,4\right)$-code, the information set  $\mathbb{A} = \left \{ 4,6,7,8 \right \}$ can thus be represented by the logical vector $\mathbf{A} = \left[ 0\,0\,0\,1\,0\,1\,1\,1 \right]$.

The code construction can be considered as an optimization problem, where the objective is to find the (sub-) optimal set of $k$ good positions in a set of indices $\left\{1,\dots,N\right\}$, as shown in (\ref{PolarCodeConstruction}). 
The reliability measure is the main difference between various code construction algorithms, e.g., upper bound on the \ac{BLER}, exact \ac{BLER} or \ac{BER} as shown in (\ref{PolarCodeConstruction}). 

\begin{equation}
\begin{aligned}
& \mathbf{A}_{\text{opt}} =\arg \underset{\mathbf{A}}{\text{      min}}
& & \mathrm{BER}(\mathbf{A}) \text{ or } \mathrm{BLER}(\mathbf{A})\\
& \text{subject to}
&& \left( \sum_{i=1}^{N} a_i \right) = k, \\
&&& a_i \in \left \{ 0,1 \right \}. \\
\end{aligned}
\label{PolarCodeConstruction}
\vspace{-0.05cm}
\end{equation}
where $\mathbf{A} = \left[ a_1,a_2,\dots,a_N \right]$, $R_c = k / N$ and $i = 1,2,\dots,N$.

The information set $\mathbb{A}$ (or, equivalently, the $\mathbf{A}$-vector) is channel dependent, meaning that it depends on the respective channel parameter (e.g., design SNR for \ac{AWGN} channel, or design $\epsilon$ for \ac{BEC}). Thus, polar codes are non-universal. However, techniques are proposed in \cite{UniversalPolarization} to devise universal polar codes. In \cite{ArikanBP}, Arıkan introduced \emph{adaptive} polar codes, or channel-tailored polar codes, in which the polar code is designed using a channel-specific condition, i.e., setting the design SNR equal to the SNR of the transmission channel. 
It is worth mentioning that the minimum distance of polar codes $d_{min}$ depends on the polar code construction (i.e., $\mathbb{A}$). Polar codes $d_{min}$ is equal to the minimum weight of the rows $\left\{\mathbf{r}_i\right\}$ in the $\mathbf{G}_N$-matrix with indices in $\mathbb{A}$ (i.e., $i\in\mathbb{A}$) \cite[Lemma~3]{Urbanke_chCsC_BP}.

Choosing the best $k$ bit positions for information transmission is even more crucial for short length polar codes. This can be attributed to the fact that the bit-channels of short length polar codes are not fully polarized, and the portion of semi-polarized bit-channels (which would be normally unfrozen) leads to high error-rates. Although efficient polar code construction only exists for the \ac{BEC} case \cite{ArikanMain}, many algorithms were devised for the \ac{AWGN} channel case. A survey on the effect of the design SNR and the effect of the specific polar code construction algorithm used on the error-rate performance of \ac{SC} decoding is presented in \cite{Vangala}.

In \cite{ArikanMain}, Arıkan uses the symmetric capacity $I(W_i)$ or the Bhattacharyya parameter $Z(W_i)$ of the virtual channel $W_i$ to assess the reliability of the bit-channels (i.e., denoted as ``conventional construction'' throughout this work). However, the Bhattacharyya parameters are preferred because of being connected with an explicit bound on the block error probability under \ac{SC} decoding. A Monte-Carlo-based polar code construction was also proposed by Arıkan in \cite{ArikanMain}.

In \cite{constructDE}, a density evolution-based polar code construction algorithm was proposed. A Gaussian approximation of the density evolution for polar code construction was proposed in \cite{constructGaussian}. In \cite{GA}, with the help of Gaussian approximation, the approximated \ac{BLER} is used to assess the reliabilities of the bit-channels instead of the upper bound on the \ac{BLER} (i.e., Bhattacharyya-based design), assuming \ac{SC} decoding.

It is important to keep in mind that polar codes that are constructed based on the mutual information or the Bhattacharyya parameters of the bit-channels, as proposed by Arıkan, are tailored to hard-output \ac{SC} decoders. Thus, they are not necessarily optimum when using other decoders such as the soft-output \ac{BP} decoder \cite{Urbanke_chCsC_BP,RMurbankePolar,BP_LLR_Siegel} or the \ac{SCL} decoder \cite{RMurbankePolar,BP_LLR_Siegel,polarDesign5G}.

In \cite{Urbanke_chCsC_BP}, the authors observed that picking the frozen bit positions according to the \ac{RM} rule  enhances the error-rate performance significantly under \ac{MAP} decoding due to the fact that the \ac{RM} rule maximizes the minimum distance $d_{min}$ of the code. An \ac{RM} code can be viewed as a polar code with a different frozen/non-frozen bit selection strategy \cite{Urbanke_chCsC_BP}, where both codes are based on the same polarization matrix $\mathbf{G}_N=\mathbf{F}^{\otimes n}$. However,  the $k$ information bit positions of  \ac{RM} codes are the positions corresponding to the $k$ row indices with the maximum weights in the $\mathbf{G}_N$-matrix. Consequently, the RM code construction phase is channel independent as it merely depends on the row weights of the $\mathbf{G}_N$-matrix. A hybrid polar and RM code construction \cite{HybridTse} results in significant error-rate improvement gains under \ac{SCL} decoding without the \ac{CRC}-aid, by improving the minimum distance of the resultant code. This underlines the benefits of improving the minimum distance of the code and also the need of an improved construction algorithm tailored to the decoder. Similarly, improved error-rate performance of multi-kernel polar codes under \ac{SCL} decoding was achieved by optimizing the distance properties of polar codes as shown in \cite{minDisLand}. A family of codes that interpolates between polar and \ac{RM} codes was introduced in \cite{RMurbankePolar}. These codes pass smoothly from a polar to an \ac{RM} code, for a fixed codelength and rate, by changing a design parameter $\alpha$. These codes provide significant error-rate performance gains under \ac{BP} and \ac{SCL} decoding.

In \cite{BP_LLR_Siegel}, an \ac{LLR}-based polar code construction is proposed, in which the \ac{LLR}s of the non-frozen bits are tracked during \ac{BP} decoding to identify weak information bit-channels, and then the information set $\mathbb{A}$ is modified by swapping these weak information bit-channels with strong frozen bit-channels. The resulting code shows an enhanced error-rate performance under both \ac{BP} and \ac{SCL} decoding due to the resultant reduction in the number of low weight codewords.

Furthermore, some work has been done to construct polar codes which are tailored to a specific decoder, e.g., polar code construction assuming \ac{SCL} decoding \cite{TUM_SCL_Construct} and assuming \ac{BP} decoding \cite{MC_BP,MC_BP_2}, where a Monte-Carlo-based construction is proposed similar to \cite{ArikanMain}.

As the output alphabet size grows exponentially with the codelength $N$, it is computationally of high complexity to precisely calculate $I(W_i)$ or $Z(W_i)$ per bit-channel. However, a quantization can be used to closely approximate them \cite{constructTalVardy}.
A recent discovery which reduces the complexity of the polar code construction is the partial order for synthesized channels \cite{UPO}, that is independent of the underlying binary-input channel.
According to \cite{Marco_Sublinear}, it suffices to compute the Bhattacharyya parameter (or any other reliability parameter) of a sub-linear number of bit-channels, if we take advantage of the partial order.
Later an heuristic closed-form algorithm called polarization weight (PW) \cite{PW} was proposed to determine the reliability of bit-channels based on their indices and a carefully chosen parameter $\beta$ \cite{BetaIngmard}, resulting in a significant complexity reduction in the polar code construction. 

To the best of our knowledge, no analytical polar code construction rule exists thus far which would be \emph{optimized} for \ac{BP} or \ac{SCL} decoding and, thus, the nature of the iterative or list decoding is not usually taken into account while designing the information set of a polar code \cite{polarDesign5G}. Thus, the problem of taking the type of decoding into consideration while constructing the polar code is, thus far, an open problem. In this work, we propose a method which always converges to a ``good enough'' solution, i.e., leads to a better error-rate performance when compared to the state-of-the-art polar code construction techniques available nowadays.

\section{Genetic Algorithm} \label{sec:genAlg}

\ac{GenAlg} was first introduced by Holland in 1975 \cite{GeneticsFirstpaper} as an efficient tool that helps in achieving (good) solutions for high-dimensional optimization problems which are computationally intractable. Beside finding (good) local minima, a well-parametrized \ac{GenAlg} is known for converging to these minima very quickly \cite{GAturbo}. Due to that merit, \ac{GenAlg} has attracted a lot of research in the \ac{AI} field leading to improved and adaptive variants of it. 

\begin{figure}[t]	
	\centering
	\resizebox{1\columnwidth}{!}{
	\includegraphics{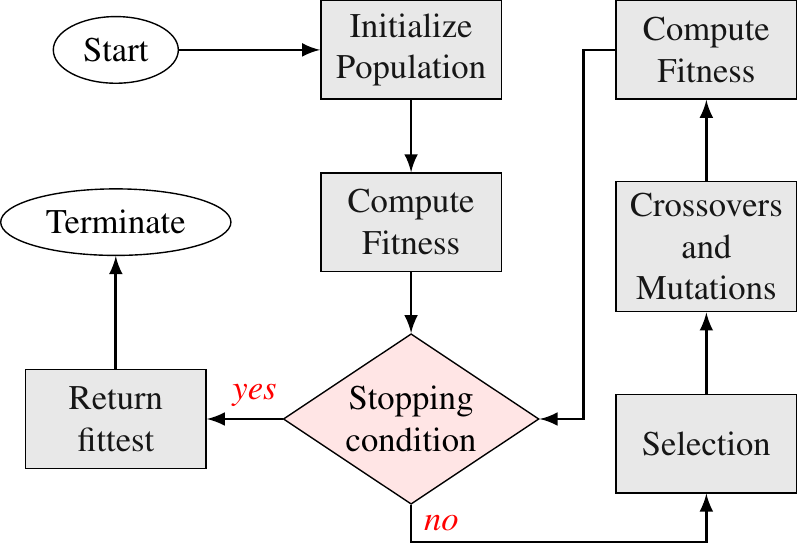}
	} 
\caption{\small An abstract view of the genetic algorithm (GenAlg).}
\label{fig:block-diagram-genAlg}  \vspace{-0.15cm}	
\end{figure}

Furthermore, researchers from other research fields (e.g., channel coding, signal processing) worked on adapting the \ac{GenAlg} to some of their specific problems that lacked enough theoretical basis for solving. This lead to improvements over the existent theoretical methods of dealing with the same problems. One of the earliest applications of \ac{GenAlg} in the field of channel coding was proposed in \cite{GAMRD} where the \ac{GenAlg} was used to find maximal distance codes. Besides, it was used in \cite{GAlinearBlockCodes} for decoding linear block codes.  

Later, it was proposed in \cite{GAturbo} to enable discarding the systematic bits at the encoder side and help reconstructing them at the decoder side using the \ac{GenAlg} and, thus, enabling rate adjustments without puncturing. Also, a \ac{GenAlg}-based \ac{LDPC} decoder was developed \cite{GAldpcdec} where the SNR value is not required at the receiver side. Furthermore, a \ac{GenAlg}-based decoder for convolutional codes was introduced in \cite{GAconv}. 

An abstract view of the \ac{GenAlg} setup is depicted in Fig.~\ref{fig:block-diagram-genAlg}. We briefly review the \ac{GenAlg}, revisit the most significant definitions relevant  to the scope of this work and refer the interested reader to \cite{GAsurvey} for more details.
\ac{GenAlg} is inspired by the natural evolution where \emph{populations} of \emph{offsprings}, resulting from \emph{parents}, keep \emph{evolving} and compete, while only the \emph{fittest} offsprings are the ones that \emph{survive}. 

\ac{GenAlg} tries to mimic the evolution of natural organisms where it typically starts with an initial population of candidate individuals and the fittest of them survive and give birth to new offsprings which represent the new population. The criterion of fitness is, therefore, critical for the \ac{GenAlg} so that the offsprings keep evolving towards the fittest. If the size of population and the number of evolution stages are sufficiently large, the \ac{GenAlg} converges to an ultimate (sub-) optimal solution.
The most important \ac{GenAlg} fundamentals are briefly introduced in the following subsections.

\subsection{Population}
The population, in the \ac{GenAlg} context, is the collection of individuals that sample the search space at any arbitrary search instance. Each individual (also called ``chromosome") is, for instance, a binary vector of bits (hence called ``genes"). These individuals represent the candidate solutions to the objective function of the optimization problem. Each of the individuals is tagged with its own fitness value according to some fitness function. One important parameter of the \ac{GenAlg} is the (allowed) population size which impacts the optimization problem significantly. The larger the size, the better solution the optimization yields and, however, the slower the convergence speed.

\subsection{Fitness function}
The fitness function, which is most commonly the objective function to be optimized, provides the baseline on which the individuals of a population are to be evaluated and, thus, allowed to survive. It provides a fitness value for each individual which serves as guideline for the \ac{GenAlg} in the direction of the optimal solution. The precise choice of the fitness function, thus, plays an important role in the quality of the final solution and the speed of converging to it. One important property of the fitness function is its speed of computation as it is extensively executed in a typical optimization problem. A more accurate estimation (or computation) of the fitness function usually enhances the quality of the final solution.

\subsection{Initialization}
\ac{GenAlg} typically starts with a randomly generated population of candidate individuals where each of them compete against each other and only the few fittest of them survive. The surviving individuals will then encounter evolutionary transformations (namely: mutations and crossovers)  to generate offsprings which would represent the new population. However, a much faster and, possibly, better convergence is often achieved by an initial population of estimates that are good enough according to their fitness values. It is worth-mentioning that a more diverse initial population can enhance the quality of the acquired solution (i.e., widened search space).

\subsection{Selection}
Selection indicates how the \ac{GenAlg} selects the parents of the next offsprings at each evolution stage, where fitter individuals are forwarded as parents for the upcoming offsprings and the weak ones perish. Many selection strategies exist \cite{GAtruncation}, where they mainly differ in the criterion of parent selection and offspring proportion dedicated for each of the selected parents. The selection strategy applied in this work is called ``Truncation selection" strategy where the population is \emph{truncated} to the fittest $T$ individuals and \emph{selected} to be the parents of the next offsprings. 

\subsection{Crossover}

Crossover is the most significant operation to which \ac{GenAlg}'s evolution towards the (sub-) optimal solution is mostly attributed. Two parent vectors among the population are subjected to crossover, by selecting a random crossover point (or, generally, several points), such that the two parent vectors exchange their genes (i.e., bits) up to the crossover point. The simplest choice of the crossover point is a single midpoint crossover. The resultant vectors are to undergo competition with other offsprings for survival (i.e., \emph{survival of the fittest}). Typically, crossover occurs at a user-defined rate called ``crossover rate" $p_c$.

\subsection{Mutation}
Mutation is the other evolutionary transformation where a small random perturbance is caused to the individual vector offsprings. This perturbance is, in its most common and simplest form, a bit flip in a random position. Mutation, thus, provides more diversity and broadens the search space. Furthermore, mutation helps in reducing the occurrence of a famous phenomenon called \emph{premature convergence}, where early convergence to a sub-optimal solution occurs due to the reduced population diversity \cite{GApreConv}. For instance, if all individuals in a population converged to a bit $0$ at position $i$ while the optimal solution has bit $1$ at this position, only mutation can find a way back towards \emph{this} optimal solution. Similar to crossover, mutation occurs at a user-defined rate called ``mutation rate" $p_m$.

\section{Genetic Algorithm-based polar code construction} \label{sec:genAlgcon}
As mentioned earlier, polar code construction can be viewed as an optimization problem (see (\ref{PolarCodeConstruction})) searching for the optimum information set $\mathbb{A}$ that has the minimum (possible) cost function. This optimization problem can be solved using \ac{GenAlg}. 
The \ac{BER} has been selected as the cost function throughout this work for optimization conducted on \ac{AWGN} channels, in order to be consistent with the results in \cite{talvardyList}.
Whereas \ac{BLER} has been selected for optimization conducted on Rayleigh fading channels, in order to be consistent with the results in \cite{TrifonovRayleigh}. Moreover, the complexity can be considered as the cost function (e.g., minimum list size in \ac{SCL} decoding or minimum number of \ac{BP} iterations, to achieve a target error-rate). All presented results throughout this work are simulated on GPUs to accelerate our error-rate simulations \cite{cammerer_HybridGPU}. Next, we discuss the polar code construction scheme based on \ac{GenAlg}, Algorithm \ref{GAalg}. The whole setup is depicted in Fig. \ref{fig:block-diagram}.

\begin{figure}[t]	
	\centering
	\resizebox{1\columnwidth}{!}{
		\includegraphics{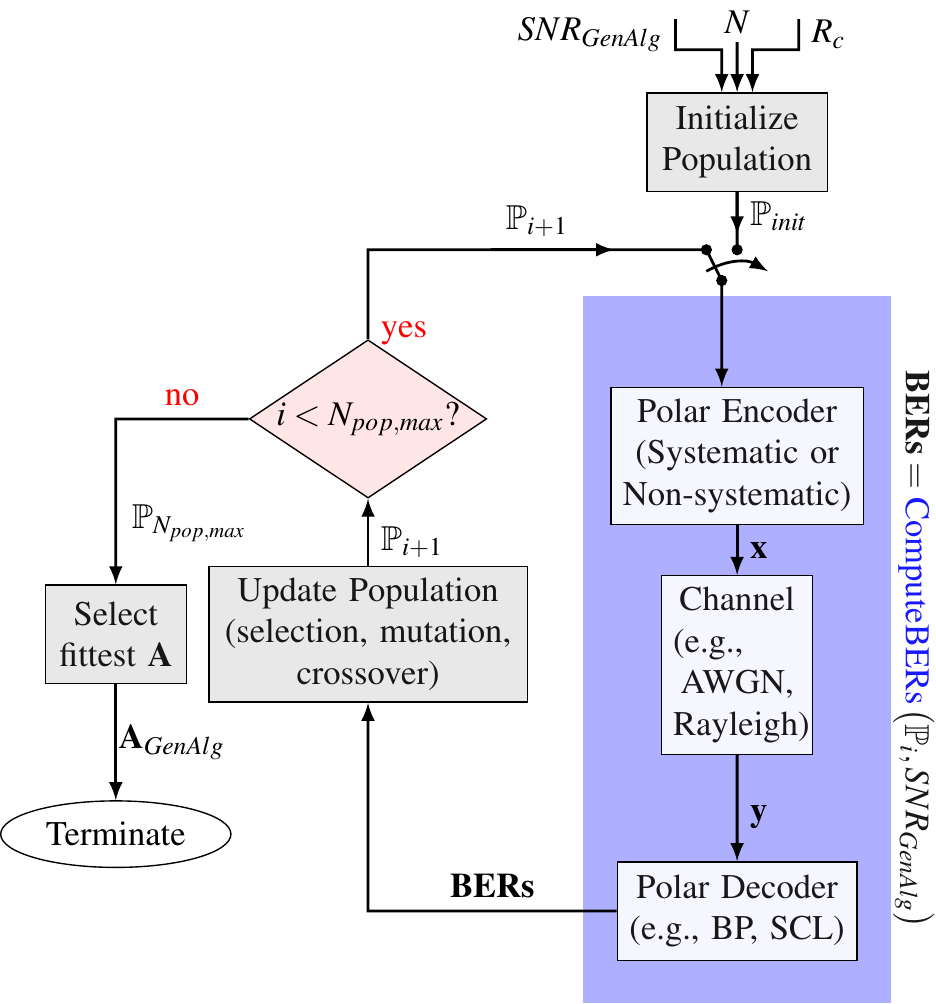}
	} 
	\vspace{-0.0cm}
	\caption{\small An abstract view of the genetic algorithm (GenAlg)-based polar code construction.}
	\label{fig:block-diagram}  
\end{figure}

\begin{algorithm}[t]
	\caption{Genetic Algorithm-based Polar Codes} \label{GAalg}	
	\begin{algorithmic}[1]
		\renewcommand{\algorithmicrequire}{\textbf{Input:}}
		\renewcommand{\algorithmicensure}{\textbf{Output:}}
		
		\Require 
		\Statex  $N$, \Comment{codelength}
		\Statex  $R_c$, \Comment{code rate}
		\Statex  $SNR_{GenAlg}$, \Comment{design SNR $\left(E_b/N_0\right)$ of the \ac{GenAlg}}
		\Statex  $N_{pop,max}$, \Comment{maximum number of populations}
		\Statex  $T$, \Comment{number of truncated parents}
						
		\Ensure  
		\Statex  $\mathbf{A}_{GenAlg}$. \Comment{optimum $\mathbf{A}$-vector}
		\Statex

   		\State $S \gets 0.5\cdot(T^2 +3T)$ 
   		\Statex \Comment{Population size, see $\left(\ref{Eq:S}\right)$}
		\State	$\mathbb{P}_{init} \gets $ \texttt{initializePopulation}$(N,R_c,SNR_{GenAlg},S)$
		\For{$i=1,2,\dots,N_{pop,max}$}
		\State	$\mathbb{P} \gets $ \texttt{updatePopulation}$(\mathbb{P},\,\mathbf{BERs},\,T)$
		\State  $\mathbf{BERs} \gets $ \texttt{computeBERs}$(\mathbb{P},\,SNR_{GenAlg})$
		\EndFor
		\State  $\mathbf{A}_{GenAlg}\gets  \texttt{select fittest } \mathbf{A} \texttt{-vector from } \mathbb{P}$
		
	\end{algorithmic}
\end{algorithm}

\subsection{Population}
In this work, the population $\mathbb{P}=\left\{\mathbf{A}_i\right\}$, for $i=1,\dots,S$, is a collection of $S$ candidate information sets represented by their binary $\mathbf{A}$-vectors, each with its own fitness value (e.g., \ac{BER} or \ac{BLER}), that represent the search space.

\subsection{Initialization}
As it turns out, \ac{GenAlg} converges faster, and probably to a better solution, if its population $\mathbb{P}_{init}$ is initialized with a sufficiently good collection of estimated $\mathbf{A}$-vectors. For that purpose, the population is initially filled with a collection of $\mathbf{A}$-vectors, all based on the Bhattacharyya construction \cite{ArikanMain} obtained for BECs with various erasure probabilities (see Algorithm \ref{initPopAlg}). Besides, we considered having the $\mathbf{A}$-vectors constructed according to \cite{HybridTse} among the initial population, in the \ac{SCL}-tailored polar code construction phase, which improved the acquired solution $\mathbf{A}_{GenAlg}$ remarkably. Furthermore, one can possibly use the codes from \cite{BP_LLR_Siegel,RMurbankePolar} in the initial population to ensure sufficient population diversity needed for quick convergence.

\begin{algorithm}[t]
	
	\caption{\texttt{initializePopulation}} \label{initPopAlg}	
	\begin{algorithmic}[1]
		\renewcommand{\algorithmicrequire}{\textbf{Input:}}
		\renewcommand{\algorithmicensure}{\textbf{Output:}}
		
		\Require 
		\Statex  $N$, \Comment{codelength}
		\Statex  $R_c$, \Comment{code rate}
		\Statex  $SNR_{GenAlg}$, \Comment{design SNR of the \ac{GenAlg}}
		\Statex  $S$, \Comment{population size}
		
		\Ensure  
		\Statex  $\mathbb{P}_{init}$. \Comment{initial population}
		\Statex

		\State $\mathbf{desSNR} \gets \{0,\cdots,5\} \text{ dB}$ 
		\Statex \Comment{design SNRs of $\mathbb{P}_{init}$ or equivalently}
		\Statex \Comment{various BEC erasure probabilities}		
		
		\State $\texttt{initialize an empty population } \mathbb{P}_{init}$

		\For{$\textbf{each } SNR \text{ in } \mathbf{desSNR}$}
		\State $\mathbf{A}  \gets $\texttt{BhattacharyyaConstruction}$(N,\, R_c,\, SNR)$
		\Statex \Comment{construction according to \cite{ArikanMain}}
		\State \texttt{add } $\mathbf{A}$ \texttt{ to }	$\mathbb{P}_{init}$

		\EndFor
		\State  $\mathbf{BERs} \gets $\texttt{computeBERs}$(\mathbb{P}_{init},\,SNR_{GenAlg})$
		\State  $\mathbb{P}_{init} \gets \texttt{select fittest } S \texttt{ } \mathbf{A} \texttt{-vectors in } \mathbb{P}_{init} $

	\end{algorithmic}
\end{algorithm}

\subsection{Fitness function}
The cost function chosen in this work is the error-rate (i.e., BER or BLER) at a user-defined design SNR $\left(SNR_{GenAlg}\right)$. The fitness function that decides the rank of each of the individual $\mathbf{A}$-vectors is, thus, selected to be the inverse of the error-rate. In other words, the $\mathbf{A}$ leading to the minimum error-rate is announced to be the optimum $\mathbf{A}$ (see Algorithm \ref{computeBERalg}). We noticed that an accurate error-rate simulation significantly improves the quality of the acquired solution $\mathbf{A}_{GenAlg}$. Alternatively, one might consider using mutual information, \ac{LLR}-reliability $\sum_{i=1}^{N}\left|LLR_i\right|$ or any other reasonable metric as the fitness function. This is left as an open research point.

\begin{algorithm}[t]
	\caption{$\texttt{computeBERs}$} \label{computeBERalg}	
	\begin{algorithmic}[1]
		\renewcommand{\algorithmicrequire}{\textbf{Input:}}
		\renewcommand{\algorithmicensure}{\textbf{Output:}}
		
		\Require 
		\Statex  $\mathbb{P}_{input}$, \Comment{input population of $\mathbf{A}$-vectors}
		\Statex  $SNR_{GenAlg}$, \Comment{design SNR of the \ac{GenAlg}}
		
		\Ensure  
		\Statex  $\mathbf{BERs}$. \Comment{\ac{BER}-vector corresponding to $\mathbb{P}_{input}$}
		\Statex
		
		\State $\texttt{initialize an empty vector } \mathbf{BERs}$
		\For {$\textbf{each } \mathbf{A} \text{ in } \mathbb{P}_{input}$}
		\State $\text{BER}  \gets \texttt{polarDecode}(\mathbf{A}, \, SNR_{GenAlg})$ 
		\Statex   \Comment{simulate a specific $\mathbf{A}$-vector under}
		\Statex   \Comment{desired decoder (e.g., \ac{BP}, \ac{SCL}, $\dots$)}
		\Statex   \Comment{over a specific channel @ $SNR_{GenAlg}$}	
		\State $\texttt{add } \text{BER} \texttt{ to }\mathbf{BERs}$
		\EndFor
		
	\end{algorithmic}
\end{algorithm}

\subsection{Mutation}
Mutation guarantees more diversity and acts against premature convergence at a certain bit position. It is, straightforwardly, a bit flip of a random position in the $\mathbf{A}$-vector representing a frozen-to-non-frozen (or a non-frozen-to-frozen) switch at that respective bit position. A mutation example is shown in Fig. \ref{fig:mutation}, where the offspring vector $\mathbf{Y}$ is the result of bit flipping the $2^{nd}$ bit of the parent vector $\mathbf{X}$. 
However, the polar code construction problem has the constraint that the number of non-frozen bit positions (i.e., number of ones) is equal to $k$ to maintain the code rate $R_c=k/N$. To restore the code rate $R_c$, one further mutation is applied to the resultant offspring vector $\mathbf{Y}$ yielding the vector $\mathbf{Z}$. Thus, the overall operation is a ``swap'' of a frozen and a non-frozen bit position. The pseudo algorithm of the \emph{mutation} (i.e., swapping) is shown in  Algorithm \ref{mut}. It can be clearly seen that for this specific problem the mutation rate $p_m$ was chosen to be 1 (i.e., one mutation always occurs for each parent per evolutionary step).

\begin{figure}[t]
	\vspace{-0.0cm}
	\captionsetup[subfigure]{position=b}
	\centering
	\begin{subfigure}{0.425\columnwidth}
		\centering 
		\includegraphics{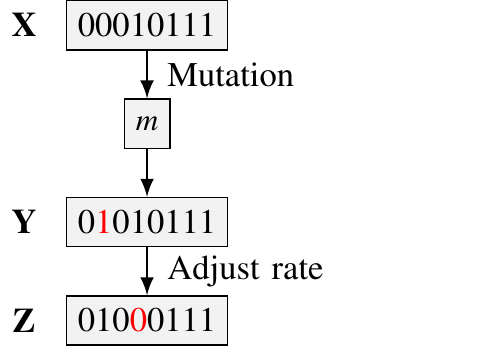}
		\vspace{-0.3cm} \caption{ Mutation or swapping}
		\label{fig:mutation}
	\end{subfigure}   \hspace{-1cm}
	\begin{subfigure}{0.425\columnwidth}
		\centering
		\includegraphics{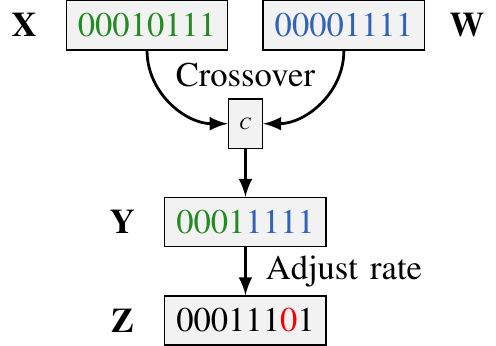}
		\vspace{-0.3cm}\caption{Crossover}
		\label{fig:crossover}
	\end{subfigure}   

	\caption{\small Examples of \ac{GenAlg}'s evolutionary transformations in the polar code construction context. Inputs $\left(\mathbf{X} \ \text{and/or} \ \mathbf{W}\right)$ and output $\left(\mathbf{Z}\right)$ satisfy the constraints in (\ref{PolarCodeConstruction}).}
	\label{fig:block-diagram-mut-cross}
\end{figure}

\begin{algorithm}[h]
	\caption{$\texttt{updatePopulation}$} \label{upd}	
	\begin{algorithmic}[1]
		\renewcommand{\algorithmicrequire}{\textbf{Input:}}
		\renewcommand{\algorithmicensure}{\textbf{Output:}}
		
		\Require 
		\Statex  $\mathbb{P}_{input}$, \Comment{input population of $\mathbf{A}$-vectors}
		\Statex  $\mathbf{BERs}$, \Comment{\ac{BER}-vector corresponding to $\mathbb{P}_{input}$}
		\Statex  $T$, \Comment{number of truncated parents}
		
		\Ensure  
		\Statex  $\mathbb{P}_{out}$. \Comment{new population after \ac{GenAlg}}
		\Statex  \Comment{transformations}
		
		\State $\texttt{initialize empty population } \mathbb{P}_{out}$
		\State  $\mathbf{newParents} \gets \texttt{fittest } T \texttt{ } \mathbf{A} \texttt{-vectors from } \mathbb{P}_{input} $	
		
		\State  $\texttt{add } \mathbf{newParents} \texttt{ to }\mathbb{P}_{out}$ 
		
		\For{$\textbf{each } \mathbf{A} \text{ in } \mathbf{newParents}$}
		\State $\mathbf{m} \gets \texttt{mutation}(\mathbf{A})$
		\State  $\text{add } \mathbf{m} \text{ to }\mathbb{P}_{out}$ 
		\EndFor 
		
		\For{ $\mathbf{each}$ pair $\mathbf{A}_1,\mathbf{A}_2$ in $\mathbf{newParents}$}
		\State $\mathbf{c} \gets \texttt{crossover}(\mathbf{A}_1,\mathbf{A}_2)$
		\State  $\text{add }\mathbf{c}\text{ to }\mathbb{P}_{out}$ 
		\EndFor 
		
	\end{algorithmic}
	
\end{algorithm}

\begin{algorithm}[t]
	\caption{$\texttt{mutation}$, see Fig. \ref{fig:mutation}} \label{mut}	
	\begin{algorithmic}[1]
		\renewcommand{\algorithmicrequire}{\textbf{Input:}}
		\renewcommand{\algorithmicensure}{\textbf{Output:}}
		
		\Require 
		\Statex  $\mathbf{A}_{input}$, \Comment{input $\mathbf{A}$-vector}

		\Ensure  
		\Statex  $\mathbf{A}_{out}$. \Comment{output $\mathbf{A}$-vector after swapping a frozen}
		\Statex
		\Comment{and a non-frozen bit position}
		\Statex
		\Comment{we assume $p_m=1$}
		\Statex
		\State $\mathbf{A}_{out} \gets \mathbf{A}_{input}$  \Comment{initialize}
		
		\State $idxs\_ones \gets \texttt{indices of 1's in }\mathbf{A}_{out}$ 
		\State $idxs\_zeros\gets \texttt{indices of 0's in }\mathbf{A}_{out}$ 
		
		\State $i \gets \texttt{random index from }idxs\_ones$
		\State $j \gets \texttt{random index from }idxs\_zeros$
		
		\State $\mathbf{A}_{out}[i] \gets 0$ \Comment{random bit flip $1$ to $0$ at position $i$}
		
		\State $\mathbf{A}_{out}[j] \gets 1$ \Comment{restore code rate $R_c$}

	\end{algorithmic}
\end{algorithm}

\subsection{Crossover}
The crossover applied throughout this work is a single midpoint crossover (see Fig. \ref{fig:crossover}), where the $1^{st}$ half of the first parent vector $\mathbf{X}$ is combined with the $2^{nd}$ half of the second parent vector $\mathbf{W}$ to generate the vector $\mathbf{Y}$. 
This often leads to a change in the number of ones in the resulting vector $\mathbf{Y}$ (i.e., remember from (\ref{PolarCodeConstruction}) that the total number of ones in the $\mathbf{A}$-vector should be equal to $k$ and thus the code rate stays constant $R_c=k/N$). To restore the code rate $R_c$, sequential bit flipping operations are applied to the resultant vector until reaching the $k$ ones in the binary $\mathbf{A}$-vector ($\mathbf{Z}$ in Fig. \ref{fig:crossover}). The pseudo algorithm of the \emph{crossover} is shown in  Algorithm \ref{cross}. Similar to mutation, the crossover rate $p_c$ was chosen to be 1 (i.e., one crossover always occurs for each pair of parents per evolutionary step). 
\begin{algorithm}
	\caption{$\texttt{crossover}$, see Fig. \ref{fig:crossover}} \label{cross}
	\begin{algorithmic}[1]
		\renewcommand{\algorithmicrequire}{\textbf{Input:}}
		\renewcommand{\algorithmicensure}{\textbf{Output:}}
		
		\Require 
		\Statex  $\mathbf{A}_{{input}_{1}}$, \Comment{$1^{st}$ input $\mathbf{A}$-vector i.e., parent 1}
		\Statex  $\mathbf{A}_{{input}_{2}}$, \Comment{$2^{nd}$ input $\mathbf{A}$-vector i.e., parent 2}

		\Ensure  
		\Statex  $\mathbf{A}_{out}$. \Comment{output $\mathbf{A}$-vector}
				\Statex
				\Comment{we assume $p_c=1$}

		\Statex
		\State $N \gets \texttt{length of }\mathbf{A}_{{input}_{1}}$ 
		\State $R_{c} \gets \frac{1}{N} \cdot \sum_{i=1}^{N}(\mathbf{A}_{{input}_{1}})$
		\State $\mathbf{A}_{out} \gets \texttt{concat}\left([\mathbf{A}_{{input}_{1}}]_1^{N/2},[\mathbf{A}_{{input}_{2}}]_{N/2+1}^{N}\right)$ 
		\Statex \Comment{concatenate two halves of}
		\Statex \Comment{$\mathbf{A}_{{input}_{1}}$ and $\mathbf{A}_{{input}_{2}}$}				
		\State $R_{c_{out}} \gets  \frac{1}{N} \cdot \sum_{i=1}^{N}(\mathbf{A}_{out})$
		\While{$R_{c_{out}} \neq R_c$}
		\State $\texttt{bit flip in }\mathbf{A}_{out}$
		\State $R_{c_{out}} \gets  \frac{1}{N} \cdot \sum_{i=1}^{N}(\mathbf{A}_{out})$
		\EndWhile
		
	\end{algorithmic}
\end{algorithm}

\subsection{Selection and population update}
In this context, by the terms \emph{selection} and \emph{population update} we mean picking the fittest $\mathbf{A}$-vectors and then applying the evolutionary transformations (namely: mutations/swapping and crossovers) in order to generate the new population. We applied the following scheme in order to generate the new population:
\begin{itemize}
	\item The fittest $T$ $\mathbf{A}$-vectors are always pushed forward as members of the new population (i.e., self-offsprings). Although this could be slowing down the convergence, this, however, ensures convergence to the (local) optimum and guarantees a monotonic behaviour of the cost function through evolving populations which facilitates observing the candidate solutions.
	\item Crossovers are applied between each pair of the  fittest $T$ $\mathbf{A}$-vectors, resulting in new ${T \choose 2}$ offsprings.
	\item Mutations (i.e., swapping operations) are applied on the  fittest $T$ $\mathbf{A}$-vectors, resulting in new $T$ mutated offsprings.
\end{itemize}
Consequently, the size of the new population $S$ is \begin{equation}S=\underset{T \ fittest}{\underbrace{T}}+\underset{crossover}{\underbrace{{T\choose 2}}} + \underset{mutation}{\underbrace{T}} = \dfrac{T^2 +3T}{2} \label{Eq:S}\end{equation}
The pseudo algorithm of the \emph{selection} and \emph{population update} scheme followed in this work is shown in  Algorithm \ref{upd}. For all simulation results using the \ac{GenAlg} as discussed next, we set $S=20$ and $T=5$.

\section{Simulation Results over \ac{AWGN} channel} \label{sec:resAWGN}

In this section, we show the results of designing polar codes using the \ac{GenAlg} method over the \ac{AWGN} channel. To be coherent with the results shown in \cite{talvardyList}, we use codes of length $N=2048$ and code rate $R_c=0.5$ in the \ac{AWGN} channel simulations. 
	 
\subsection{\ac{SC} decoder}

All of the known polar code construction (i.e., frozen bit position selection) algorithms assume that the \ac{SC} decoder is used. Thus, they all are tailored to hard-output \ac{SC} decoding. As shown in Fig. \ref{fig:SC_comp}, all construction algorithms yield codes having similar error-rate performance  over the \ac{AWGN} channel, including the \ac{GenAlg}-based construction. A similar independent observation supporting our results was presented in \cite{Vangala}.

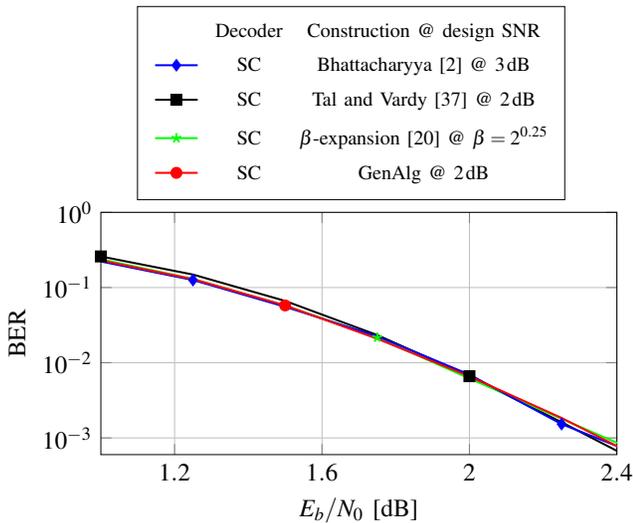
\begin{figure}[t]
	\centering
	\resizebox{0.975\columnwidth}{!}{\begin{tikzpicture}

\begin{axis}[
width=\linewidth,
height=5cm,
xmin=1,
xmax=2.4,
xlabel={$E_b/N_0$ [dB]},
xmajorgrids,
ymode=log,
ymin=0.0006,
ymax=1,
yminorticks=false,
ymajorgrids,
ylabel={BER},
xtick={1.2,1.6,2,2.4},
axis background/.style={fill=white},
]

\addplot [color=black,solid, mark=square*, thick,mark repeat=4,mark phase=0,]
  table[row sep=crcr]{
1	0.2591959375\\
1.25	0.1487934375\\
1.5	0.066401494140625\\
1.75	0.02332587890625\\
2	0.00662234375\\
2.25	0.00161689453125\\
2.5	0.000375185546875\\
2.75	6.9482421875e-05\\
};
  \label{plot:SC_TalVardy}

\addplot [color=green,solid, mark=star, thick,mark repeat=4,mark phase=4,]
  table[row sep=crcr]{
1	0.23759013671875\\
1.25	0.13016201171875\\
1.5	0.05765400390625\\
1.75	0.02164814453125\\
2	0.00618310546875\\
2.25	0.00179833984375\\
2.5	0.00052763671875\\
2.75	0.000183984375\\
};
  \label{plot:SC_Beta}

\addplot [color=blue,solid, mark=diamond*, thick,mark repeat=4,mark phase=2,]
  table[row sep=crcr]{
1	0.22210654296875\\
1.25	0.12587255859375\\
1.5	0.05551962890625\\
1.75	0.0224416015625\\
2	0.00695966796875\\
2.25	0.00152734375\\
2.5	0.00049931640625\\
2.75	0.00014560546875\\
};
  \label{plot:SC_Bha}

\addplot [color=red,solid, mark=*, thick,mark repeat=4,mark phase=3,]
  table[row sep=crcr]{
1	0.229691083984375\\
1.25	0.1293030078125\\
1.5	0.05779046875\\
1.75	0.020831005859375\\
2	0.006584833984375\\
2.25	0.001845673828125\\
2.5	0.000438193359375\\
2.75	0.000104580078125\\
};
  \label{plot:SC_GenAlg}

\coordinate (legend) at (axis description cs:0.9,1.05);
\end{axis}

    \matrix [
    draw,
    matrix of nodes,
    anchor=south east,
    font=\footnotesize,
    ] at (legend) {
    	& Decoder & Construction @ design SNR \\
   	\ref{plot:SC_Bha} & SC   & Bhattacharyya \cite{ArikanMain} @ $\unit[3]{dB}$   \\
    	\ref{plot:SC_TalVardy} & SC   & Tal and Vardy \cite{constructTalVardy} @ $\unit[2]{dB}$   \\    	
    	\ref{plot:SC_Beta} & SC   & $\beta$-expansion \cite{BetaIngmard} @ $\beta = 2^{0.25}$ \\
    	\ref{plot:SC_GenAlg} & SC  & GenAlg @ $\unit[2]{dB}$   \\
    };
    
\end{tikzpicture}}
		\caption{\small BER performance of the \ac{GenAlg}-based $\mathcal{P}$(2048,1024)-code under \ac{SC} decoding over the \ac{AWGN} channel and no CRC is used.}
	\label{fig:SC_comp}
\end{figure}

\subsection{\ac{BP} decoder}

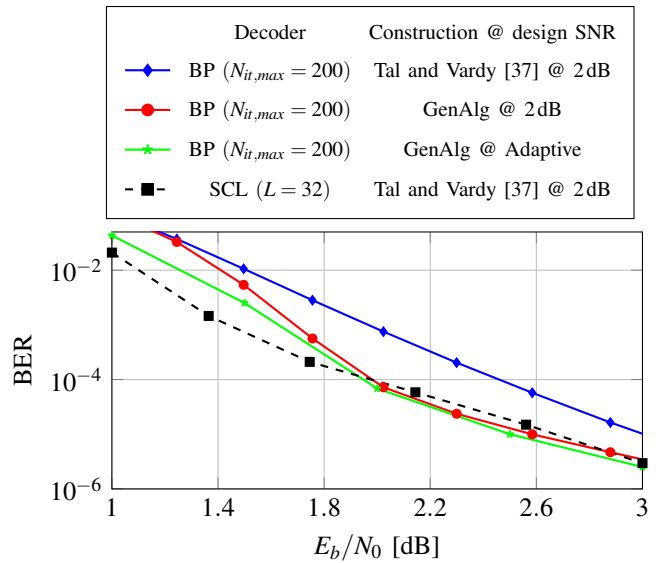
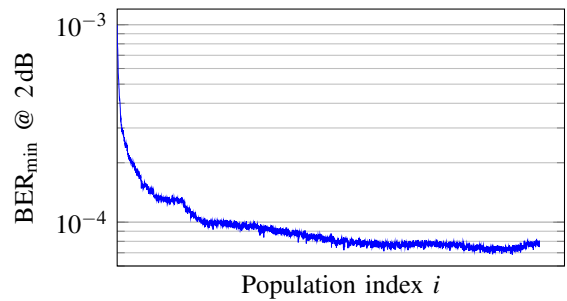
\begin{figure}[t]
	\centering
	\begin{subfigure}[t]{0.975\columnwidth}
		\begin{tikzpicture}
\begin{axis}[
width=\linewidth,
height=5cm,
xmajorgrids,
yminorticks=false,
ymajorgrids,
legend columns=1,
xlabel={$E_b/N_0$ [dB]},
ylabel={BER},
ymode=log,
mark size=1.5pt,
xmin=1,
xmax=3.0,
xtick={1,1.4,1.8,2.2,2.6,3},
ytick={1e-2, 1e-4, 1e-6},
ymin=1e-6,
ymax=5e-2,
mark options={solid}
]

\addplot [color=blue,solid, mark=diamond*, thick] 
table[row sep=crcr]{
	1.00000000000000	0.112095924497848\\
	1.24479414452514	0.0371581009546795\\
	1.49668788567703	0.0105970127808852\\
	1.75610536460180	0.00282372360618388\\
	2.02350990413076	0.000753127703832328\\
	2.29940898861021	0.000203934403722672\\
	2.58436006091173	5.73887868431984e-05\\
	2.87897730301695	1.64008483608058e-05\\
	3.18393960749088	4.96658765605456e-06\\
	3.50000000000000	1.23719593571999e-06\\
};
\label{plot:BP_TalVardy}

\addplot [color=red,solid, mark=*, thick,mark repeat=1,mark phase=0] 
table[row sep=crcr]{
	1.00000000000000	0.113772346124731\\
	1.24479414452514	0.0327804587196862\\
	1.49668788567703	0.00540290682357529\\
	1.75610536460180	0.000563304083445157\\
	2.02350990413076	7.21655672675462e-05\\
	2.29940898861021	2.36728065841675e-05\\
	2.58436006091173	9.96102220692113e-06\\
	2.87897730301695	4.68300172459776e-06\\
	3.18393960749088	2.18632506414491e-06\\
	3.50000000000000	1.06981733722886e-06\\
};
\label{plot:GenAlg_BP}

\addplot [color=green,solid, mark=star, thick,mark repeat=1,mark phase=0] 
table[row sep=crcr]{
	1	                0.0431934703480114\\
	1.50000000000000	0.00251187847336803\\
	2	                6.85561775077723e-05\\
	2.50000000000000	1.00163561617740e-05\\
	3                   2.51346902209293e-06\\ 	
	};
\label{plot:Adaptive_GenAlg_BP}

\addplot [color=black,dashed, mark=square*, thick] 
table[row sep=crcr]{
	1.00000000000000	0.0209942992156605\\
	1.36484480920620	0.00145650969990947\\
	1.74568917910050	0.000210496604956101\\
	2.14400098577864	5.84599609375000e-05\\
	2.56145986298185	1.49636509258887e-05\\
	3.00000000000000	2.95790542360132e-06\\
};
\label{plot:SCL_TalVardy}

    \coordinate (legend) at (axis description cs:0.985,1.05);
\end{axis}

    \matrix [
    draw,
    matrix of nodes,
    anchor=south east,
    font=\footnotesize,
    mark options={solid}
    ] at (legend) {
    	& Decoder & Construction @ design SNR \\
    	\ref{plot:BP_TalVardy} & BP $\left(N_{it,max}=200\right)$   & Tal and Vardy \cite{constructTalVardy} @ $\unit[2]{dB}$   \\
    	\ref{plot:GenAlg_BP} & BP $\left(N_{it,max}=200\right)$   & GenAlg @ $\unit[2]{dB}$   \\
    	\ref{plot:Adaptive_GenAlg_BP} & BP $\left(N_{it,max}=200\right)$   & GenAlg @ Adaptive   \\
    	\ref{plot:SCL_TalVardy} & SCL $\left(L=32\right)$   & Tal and Vardy \cite{constructTalVardy} @ $\unit[2]{dB}$   \\    	
    };

\end{tikzpicture}
		\vspace{-0.5cm}		
		\caption{\small BER performance}
		\label{fig:BP_comp}
	\end{subfigure}
	
	\vspace{0.4cm}	
	
	\begin{subfigure}{0.85\columnwidth}
		
		\includegraphics{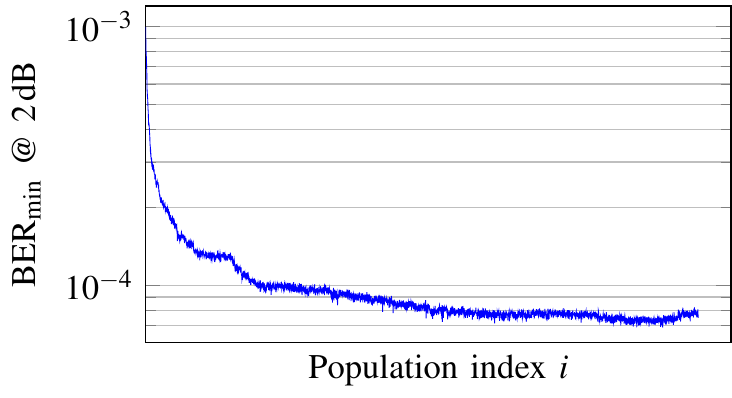}
		\caption{\small Evolution of the BER at $SNR_{GenAlg}$ $\left({E_b}/{N_0}\right) = \unit[2]{dB}$}		
		\label{fig:BP_epochs}
	\end{subfigure}
	\caption{\small GenAlg-based construction of a $\mathcal{P}$(2048,1024)-code under \ac{BP} $\left(N_{it,max}=200\right)$ decoding over the \ac{AWGN} channel and no CRC is used.}	
\end{figure}

We use the \ac{GenAlg} to design polar codes tailored to \ac{BP} decoding over \ac{AWGN} channel. Fig. \ref{fig:BP_comp} shows a \ac{BER} comparison between a code constructed via the method proposed in \cite{constructTalVardy} at a design SNR $\left({E_b}/{N_0}\right) = \unit[2]{dB}$ and a code constructed using our proposed \ac{GenAlg} at $SNR_{GenAlg}$ $\left({E_b}/{N_0}\right) = \unit[2]{dB}$ under BP $\left(N_{it,max}=200\right)$ decoding. The \ac{GenAlg}-based construction yields a $\unit[0.5]{dB}$ net coding gain at BER of $10^{-4}$ when compared to the construction proposed in \cite{constructTalVardy}. It is worth mentioning that the only difference between the two codes is the selection of the frozen/non-frozen bit positions, i.e., both use exactly the same decoder. Fig. \ref{fig:BP_epochs} shows the evolution (or enhancement) of the BER per population (i.e., we plot the minimum BER per population). Note that minor fluctuations in the \acp{BER} per population index can be explained by the Monte-Carlo output.

The $d_{min}$ of the \ac{BP}-tailored polar code using \ac{GenAlg} was found to be $8$, while the $d_{min}$ of the polar code constructed via \cite{constructTalVardy} is $16$.
This is due to the fact that the performance of a linear code under iterative decoding is dominated by the structure of the stopping sets in the Tanner graph of the code and not its $d_{min}$ \cite{vardyStoppingDistance}. 
Thus, $d_{min}$ is not the only parameter to be maximized in order to design linear codes tailored to iterative decoding. A fact supporting this claim is that the \ac{RM} code has a larger (maximum) $d_{min}$ but worse error-rate performance under iterative decoding when compared to polar codes \cite{ArikanBP}.

In the following, we assess the robustness of the GenAlg-based polar code design against inaccuracy in SNR estimation (i.e., having a mismatch between the design SNR and the operating SNR). We use the \ac{GenAlg} to design an ``adaptive'' polar code, in which the $\mathbf{A}$-vector is channel-tailored, or SNR $\left({E_b}/{N_0}\right)$-tailored (i.e., the frozen set is optimized separately for each value of the SNR of the transmission channel). The BER performance of this GenAlg-based adaptive polar code (\ref{plot:Adaptive_GenAlg_BP}) is very close to the polar code designed by the \ac{GenAlg} at $\unit[2]{dB}$ (\ref{plot:GenAlg_BP}) in the high SNR region as shown in Fig. \ref{fig:BP_comp}. This means that \ac{GenAlg} at a specific $SNR_{GenAlg}$ yields a polar code which might be good enough for a certain wider range of SNRs. 

Note that the BER performance of the \ac{BP}-tailored polar code (\ref{plot:GenAlg_BP}) is very close to the performance of a polar code constructed via the method proposed in \cite{constructTalVardy} under \ac{SCL} $\left(L=32\right)$ decoding (\ref{plot:SCL_TalVardy}), as also shown in Fig. \ref{fig:BP_comp}.
	
\subsection{\ac{SCL} decoder}

\begin{figure}[t]
	\centering
	\resizebox{0.975\columnwidth}{!}{\begin{tikzpicture}
\begin{axis}[
width=\linewidth,
height=5cm,
xmajorgrids,
yminorticks=false,
ymajorgrids,
xlabel={$E_b/N_0$ [dB]},
ylabel={BER},
ymode=log,
mark size=1.5pt,
xmin=1,
xmax=3,
ytick={1e-2, 1e-4, 1e-6},
xtick={1,1.4,1.8,2.2,2.6,3},
ymin=1e-6,
ymax=4e-2,
mark options={solid}
]

\addplot [color=blue,solid, mark=diamond*, thick]    
table[row sep=crcr]{
	1.00000000000000	0.0209942992156605\\
	1.36484480920620	0.00145650969990947\\
	1.74568917910050	0.000210496604956101\\
	2.14400098577864	5.84599609375000e-05\\
	2.56145986298185	1.49636509258887e-05\\
	3.00000000000000	2.95790542360132e-06\\
};
\label{plot:SCL_TalVardy2}

\addplot [color=green,solid, mark=*, thick] 
table[row sep=crcr]{
1.00000000000000	0.0474210698711490\\
1.36484480920620	0.00478821719327529\\
1.74568917910050	0.000186579417555733\\
2.14400098577864	2.92862819854588e-06\\
2.56145986298185	3.48079932894319e-08\\
};
\label{plot:SCL_RM_Polar}

\addplot [color=red,solid, mark=square, thick,mark repeat=1,mark phase=0] 
table[row sep=crcr]{
	1.00000000000000	0.0382738373324745\\
	1.48561678411588	0.000715864937622712\\
	2.00000000000000	1.83336318835634e-06\\
	2.20000000000000	1.94693505555143e-07\\
	};
\label{plot:GenAlg_SCL}

  \addplot [color=black,solid, mark=star, thick, dashed,mark repeat=1,mark phase=2]
  table[row sep=crcr]{
  	1	  3e-2\\
  	1.25  6e-3\\
  	1.5	  6e-4\\
  	1.75  4e-5\\
  	2	  2e-6\\
  	2.2	  4.73816037161570e-08\\ 
  };
  \label{plot:SCL_CRC_TalVardy}

\coordinate (legend) at (axis description cs:1,1.05);
\end{axis}

    \matrix [
    draw,
    matrix of nodes,
    anchor=south east,
    font=\footnotesize,
    mark options={solid}
    ] at (legend) {
    	& Decoder & Construction @ design SNR \\
    	\ref{plot:SCL_TalVardy2} & SCL $\left(L=32\right)$   & Tal and Vardy \cite{constructTalVardy} @ $\unit[2]{dB}$   \\    	
    	\ref{plot:SCL_RM_Polar} & SCL $\left(L=32\right)$   & RM-Polar \cite{HybridTse} @ $\unit[2]{dB}$   \\    	
    	\ref{plot:GenAlg_SCL} & SCL $\left(L=32\right)$   & GenAlg @ $\unit[2]{dB}$   \\
    	\ref{plot:SCL_CRC_TalVardy} & SCL+CRC-16 $\left(L=32\right)$   & Tal and Vardy \cite{constructTalVardy} @ $\unit[2]{dB}$   \\
    };

\end{tikzpicture}}
	\caption{\small BER performance of the \ac{GenAlg}-based $\mathcal{P}$(2048,1024)-code under \ac{SCL} decoding over the \ac{AWGN} channel. The CRC-aided polar code (\ref{plot:SCL_CRC_TalVardy}): $N=2048$, $k=1024$, $r=16$ and, thus, the polar code is a $\mathcal{P}$(2048,1040)-code.}		
	\label{fig:SCL_comp}
	\vspace{-0.5cm}
\end{figure}
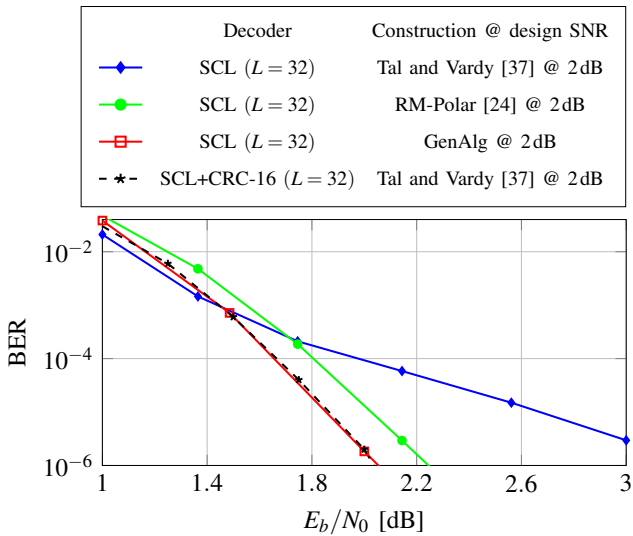

Next, the \ac{GenAlg} is applied to construct polar codes tailored to \ac{SCL} $\left(L=32\right)$ over the \ac{AWGN} channel. Fig. \ref{fig:SCL_comp} shows a \ac{BER} comparison between a code constructed via the method proposed in \cite{constructTalVardy} at SNR $\left({E_b}/{N_0}\right) = \unit[2]{dB}$ and a code constructed using our proposed \ac{GenAlg} at $SNR_{GenAlg}$ $\left({E_b}/{N_0}\right) = \unit[2]{dB}$ under \ac{SCL} decoding with list size $L=32$. The \ac{GenAlg}-based construction shows significant performance improvements, yielding a $\unit[1]{dB}$ net coding gain at BER of $10^{-6}$ when compared to the construction proposed in \cite{constructTalVardy}, where again the only difference between the two codes is the frozen/non-frozen bit positions.

The \ac{GenAlg}-optimized polar code without \ac{CRC}-aid under \ac{SCL} decoding (\ref{plot:GenAlg_SCL}) performs equally well as the \ac{CRC}-aided polar code under CRC-aided SCL decoding (\ref{plot:SCL_CRC_TalVardy}), with the same list size $L=32$ and the same code rate $R_c=0.5$, as shown in Fig. \ref{fig:SCL_comp}. 
The $d_{min}$ of the \ac{SCL}-tailored polar code using \ac{GenAlg} is $16$ which is exactly the same as the $d_{min}$ of the polar code constructed via \cite{constructTalVardy}. However, the RM-Polar code (\ref{plot:SCL_RM_Polar}) has a $d_{min}=32$.
Thus, $d_{min}$ is not the only parameter to be maximized in order to design polar codes tailored to \ac{SCL} decoding. Note that maximizing $d_{min}$ will lead to an \ac{RM} code.

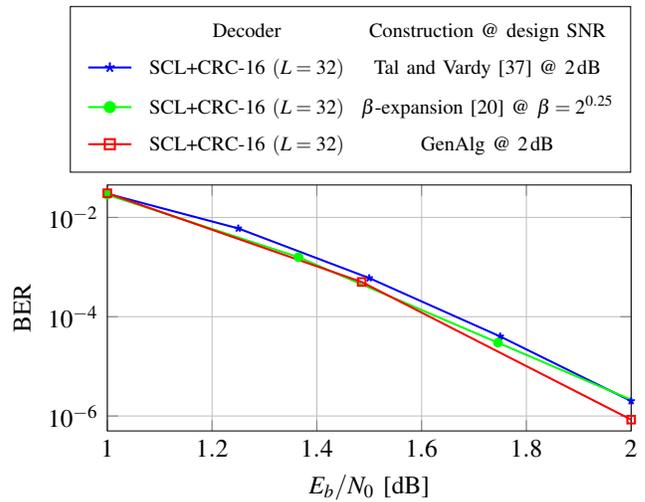
\begin{figure}[t]
	\centering
	\resizebox{0.975\columnwidth}{!}{\begin{tikzpicture}
\begin{axis}[
width=\linewidth,
height=5cm,
xmajorgrids,
yminorticks=false,
ymajorgrids,
xlabel={$E_b/N_0$ [dB]},
ylabel={BER},
ymode=log,
mark size=1.5pt,
xmin=1,
xmax=2,
ymin=5e-7,
ymax=4.5e-2,
]

  \addplot [color=blue,solid, mark=star, thick] 
  table[row sep=crcr]{
  	1	  3e-2\\
  	1.25  6e-3\\
  	1.5	  6e-4\\
  	1.75  4e-5\\
  	2	  2e-6\\
    };
      \label{plot:SCL_CRC_TalVardy2}

  \addplot [color=green,solid, mark=*, thick]
  table[row sep=crcr]{
  	1.00000000000000	0.0287536917598999\\
  	1.36484480920620	0.00158135854812568\\
  	1.74568917910050	2.99422631524094e-05\\
  	2.14400098577864	4.73937010499176e-07\\
  };
  \label{plot:SCL_CRC_Beta_AWGNC}
  
\addplot [color=red,solid,mark=square,thick]
table[row sep=crcr]{
	1	0.0307437754264963\\
	1.48561678411588	0.000500881611373141\\
	2	8.43812031333419e-07\\
};
      \label{plot:GenAlg_SCL_CRC}
      
 \coordinate (legend) at (axis description cs:1,1.05);   
\end{axis}

\matrix [
draw,
matrix of nodes,
anchor=south east,
font=\footnotesize,
] at (legend) {
	& Decoder & Construction @ design SNR \\
	\ref{plot:SCL_CRC_TalVardy2} & SCL+CRC-16 $\left(L=32\right)$   & Tal and Vardy \cite{constructTalVardy} @ $\unit[2]{dB}$   \\
	\ref{plot:SCL_CRC_Beta_AWGNC} & SCL+CRC-16 $\left(L=32\right)$   &  $\beta$-expansion \cite{BetaIngmard} @ $\beta=2^{0.25}$  \\	
	\ref{plot:GenAlg_SCL_CRC} & SCL+CRC-16 $\left(L=32\right)$   & GenAlg @ $\unit[2]{dB}$   \\	
};

\end{tikzpicture}}
	\caption{\small BER performance of the \ac{GenAlg}-based $\mathcal{P}$(2048,1024)-code under CRC-aided SCL decoding over the \ac{AWGN} channel.
		The CRC-aided polar code: $N=2048$, $k=1024$, $r=16$ and, thus, the polar code is a $\mathcal{P}$(2048,1040)-code.
	}
	\label{fig:SCL_CRC_comp}
	\vspace{-0.4cm}
\end{figure}

 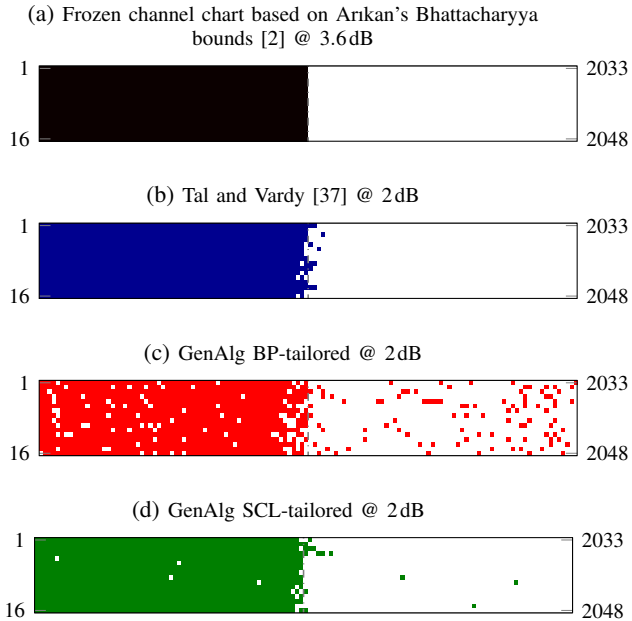
\begin{figure}[t]
 	\captionsetup[subfigure]{position=b}
 	\centering
 	
 	\begin{subfigure}{0.85\columnwidth}
 		\centering
 		\captionsetup{justification=centering}
 		\caption{\footnotesize Frozen channel chart based on Arıkan's Bhattacharyya bounds \cite{ArikanMain} @ $\unit[3.6]{dB}$}
 		\begin{tikzpicture}
 		\begin{axis}[
 		width=0.95\linewidth,
 		height=1cm,
 		scale only axis,
 		axis y line*=right,
 		xmin=0.5,
 		xmax=128.5,
 		yticklabels={2033,2048},ytick = {16,1},
 		xtick =\empty,
	    tick label style = {font = \footnotesize}, 		
 		ymin=0.5,
 		ymax=16.5,
 		axis background/.style={fill=white},
 		]
 		\end{axis}
 		\begin{axis}[
 		width=0.95\linewidth,
 		height=1cm,
 		scale only axis,
 		axis on top,
 		xmin=0.5,
 		xmax=128.5,
 		xtick={\empty},
 		y dir=reverse,
 		ymin=0.5,
 		ymax=16.5,
 		ytick={ 1, 16},
	    tick label style = {font = \footnotesize}, 	 		
 		axis background/.style={fill=white},
 		]
 		\addplot [forget plot] graphics [xmin=0.5, xmax=128.5, ymin=0.5, ymax=16.5] {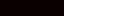};
 		\addplot +[mark=none,gray,dash dot,semithick] coordinates {(64.5, -1) (64.5, 18)};
 		\end{axis}
 		\end{tikzpicture}
 		\label{fig:Bha-hist}
 	\end{subfigure}
 	
 	\vspace{-0.25cm}
 	
 	\begin{subfigure}{0.85\columnwidth}  
 		\centering 
 		\caption{\footnotesize Tal and Vardy \cite{constructTalVardy} @ $\unit[2]{dB}$}   
 		\begin{tikzpicture}
 		\begin{axis}[
 		width=0.95\linewidth,
 		height=1cm,
 		scale only axis,
 		axis y line*=right,
 		xmin=0.5,
 		xmax=128.5,
 		yticklabels={2033,2048},ytick = {16,1},
 		xtick =\empty,
	    tick label style = {font = \footnotesize}, 	 		
 		ymin=0.5,
 		ymax=16.5,
 		axis background/.style={fill=white},
 		]
 		\end{axis}
 		\begin{axis}[
 		width=0.95\linewidth,
 		height=1cm,
 		scale only axis,
 		axis on top,
 		xmin=0.5,
 		xmax=128.5,
 		xtick={\empty},
 		y dir=reverse,
 		ymin=0.5,
 		ymax=16.5,
 		ytick={ 1, 16},
	    tick label style = {font = \footnotesize}, 	 		
 		axis background/.style={fill=white},
 		title style={font=\bfseries},
 		]
 		\addplot [forget plot] graphics [xmin=0.5, xmax=128.5, ymin=0.5, ymax=16.5] {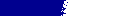};
 		\addplot +[mark=none,gray,dash dot,semithick] coordinates {(64.5, -1) (64.5, 18)};
 		\end{axis}
 		\end{tikzpicture}
 		\label{fig:TalVardy-hist}
 	\end{subfigure}
 	
 	\vspace{-0.25cm} 	
 	
 	\begin{subfigure}{0.85\columnwidth}   
 		\centering
 		\caption{\footnotesize GenAlg BP-tailored @ $\unit[2]{dB}$}  
 		\begin{tikzpicture}
 		\begin{axis}[
 		width=0.95\linewidth,
 		height=1cm,
 		scale only axis,
 		axis y line*=right,
 		xmin=0.5,
 		xmax=128.5,
 		yticklabels={2033,2048},ytick = {16,1},
 		xtick =\empty,
	    tick label style = {font = \footnotesize}, 	 		
 		ymin=0.5,
 		ymax=16.5,
 		axis background/.style={fill=white},
 		]
 		\end{axis}
 		\begin{axis}[
 		width=0.95\linewidth,
 		height=1cm,
 		scale only axis,
 		axis on top,
 		xmin=0.5,
 		xmax=128.5,
 		xtick={\empty},
 		y dir=reverse,
 		ymin=0.5,
 		ymax=16.5,
 		ytick={ 1, 16},
	    tick label style = {font = \footnotesize}, 	 		
 		axis background/.style={fill=white},
 		title style={font=\bfseries},
 		]
 		\addplot [forget plot] graphics [xmin=0.5, xmax=128.5, ymin=0.5, ymax=16.5] {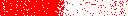};
 		\addplot +[mark=none,gray,dash dot,semithick] coordinates {(64.5, -1) (64.5, 18)};
 		\end{axis}
 		\end{tikzpicture}
 		\label{fig:GenAlg-BP-hist}
 	\end{subfigure}

 	\vspace{-0.25cm}
 	 	
 	\begin{subfigure}{0.85\columnwidth}   
 		\centering 
 		\caption{\footnotesize GenAlg SCL-tailored @ $\unit[2]{dB}$}     
 		\begin{tikzpicture}
 		\begin{axis}[
 		width=0.95\linewidth,
 		height=1cm,
 		scale only axis,
 		axis y line*=right,
 		xmin=0.5,
 		xmax=128.5,
 		yticklabels={2033,2048},ytick = {16,1},
 		xtick =\empty,
	    tick label style = {font = \footnotesize}, 	 		
 		ymin=0.5,
 		ymax=16.5,
 		axis background/.style={fill=white},
 		]
 		\end{axis}
 		\begin{axis}[
 		width=0.95\linewidth,
 		height=1cm,
 		scale only axis,
 		axis on top,
 		xmin=0.5,
 		xmax=128.5,
 		xtick={\empty},
 		y dir=reverse,
 		ymin=0.5,
 		ymax=16.5,
 		ytick={ 1, 16},
	    tick label style = {font = \footnotesize}, 	 		
 		axis background/.style={fill=white},
 		title style={font=\bfseries},
 		]
 		\addplot [forget plot] graphics [xmin=0.5, xmax=128.5, ymin=0.5, ymax=16.5] {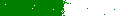};
 		\addplot +[mark=none,gray,dash dot,semithick] coordinates {(64.5, -1) (64.5, 18)};
 		\end{axis}
 		\end{tikzpicture}
 		\label{fig:GenAlg-SCL-hist}
 	\end{subfigure}
 	 	\vspace{-0.5cm}
 	\caption{\small Frozen channel chart (i.e., frozen bit position pattern) of a $\mathcal{P}$(2048,1024)-code over the AWGN channel with different polar code construction algorithms. 
 		The $2048$ bit positions are plotted over a $16 \times 128$ matrix. 
 		Note that the bit-channels are sorted with decreasing Bhattacharyya parameter value. 
 		White: non-frozen; colored: frozen.} 
 	\label{fig:hist}
 	\vspace{-0.6cm}
 \end{figure}
	
\subsection{CRC-aided SCL decoder}
In this section, we use the \ac{GenAlg} to design polar codes tailored to CRC-aided SCL decoding over \ac{AWGN} channel. Fig. \ref{fig:SCL_CRC_comp} shows a \ac{BER} comparison between a code constructed via the method proposed in \cite{constructTalVardy} at SNR $\left({E_b}/{N_0}\right) = \unit[2]{dB}$ and a code constructed using our proposed \ac{GenAlg} at $SNR_{GenAlg}$ $\left({E_b}/{N_0}\right) = \unit[2]{dB}$ under CRC-aided SCL decoding with list size $L=32$. The \ac{GenAlg}-based construction yields a slight improvement in terms of BER performance when compared to the construction proposed in \cite{constructTalVardy}. We want to emphasize that all \ac{BER} simulations are performed in the same decoding framework and, thus, the gains are \emph{real} gains caused by the code design.
	
\subsection{$\mathbf{A}$-vector Analysis}

The distribution of frozen bit-channels over the $N$ synthesized virtual channels is shown in the ``Frozen channel chart'' in Fig. \ref{fig:hist}. Every square (or pixel) in Fig. \ref{fig:hist} corresponds to a bit-channel (i.e., we show $N=2048$ pixels per sub-figure). If a pixel is white, then the corresponding bit-channel is used for information transmission (i.e., included in the information set). If a pixel is colored, then the corresponding bit-channel is a frozen bit position (i.e., included in the frozen set). In Fig. \ref{fig:Bha-hist}, we show the result of the Bhattacharyya-based construction at a specific channel parameter. For the sake of reproducibility, the frozen/non-frozen set constructed using the Bhattacharyya bounds \cite{ArikanMain} at design SNR $\unit[3.6]{dB}$ 
is equivalent to the frozen/non-frozen set constructed using the Bhattacharyya parameter for a BEC with $\epsilon=0.32$. The $N-k$ bit-channels with the highest Bhattacharyya value are set to frozen (i.e., black pixel). The $k$ bit-channels with the smallest Bhattacharyya value are set to non-frozen (i.e., white pixel). Then the bit-channels are sorted according to their Bhattacharyya value in descending order (i.e., the $N-k$ frozen bit-channels followed by the $k$ non-frozen bit-channels). And we use this ordering to represent the other three constructions (Fig. \ref{fig:TalVardy-hist}, \ref{fig:GenAlg-BP-hist} and \ref{fig:GenAlg-SCL-hist}), which clearly highlights the differences between the output of different construction algorithms.

The conventional construction techniques (e.g., \cite{ArikanMain,constructTalVardy}) assume that an SC decoder is used. Thus, these construction techniques yield very similar $\mathbf{A}$-vectors having a very similar distribution as shown in Fig. \ref{fig:Bha-hist} and \ref{fig:TalVardy-hist}. Our proposed GenAlg-based construction tailors the $\mathbf{A}$-vector (or the code) to a specific decoder (e.g., BP or SCL decoder). The BP decoder has an iterative (i.e., non-sequential) decoding nature when compared to SC decoding. Thus, the $\mathbf{A}$-vector tailored to BP decoding has a much different frozen bit-channel distribution as shown in Fig. \ref{fig:GenAlg-BP-hist}. Note that the $\mathbf{A}$-vector tailored to BP decoding contains a significant number of non-frozen bits in the first half part of the code, which is not the case in the  conventional code construction algorithms assuming \ac{SC} decoding. Similarly, the GenAlg-based construction tailored to SCL decoding should take the list decoding nature into consideration and, thus, a different frozen bit-channel distribution as shown in Fig. \ref{fig:GenAlg-SCL-hist}. By comparing Fig. \ref{fig:Bha-hist} with Fig. \ref{fig:GenAlg-BP-hist} and \ref{fig:GenAlg-SCL-hist}, in a pixel-by-pixel manner, one can see the differences in the frozen/non-frozen sets due to the assumption of a different decoder other than the SC decoder (i.e., BP and SCL) while constructing the code.

Furthermore, Table \ref{tab:decoder-tailored} investigates the effect of a mismatch between the polar code design and the polar decoder used. 
It is shown that this mismatch can lead to a rather high error-rate. 
Mismatch in this context means that a polar code which is designed tailored to decoder X is decoded with another different decoder Y.
Table \ref{tab:decoder-tailored} shows that 
\begin{itemize}
	\item for SC decoding, the best construction method is \cite{constructTalVardy}
	\item for BP decoding, the best construction method is our proposed GenAlg
	\item for SCL decoding, the best construction method is our proposed GenAlg.
\end{itemize}
Thus, the best case (i.e., achieving a target BER with the minimum SNR) is to have a polar code tailored to the used decoder. It should be emphasized that no additional CRC was used for the results in Table \ref{tab:decoder-tailored}.

\begin{table}[h]
	\begin{center}
		\caption{\small Illustration of polar design and decoder architecture mismatch by evaluating the minimum $E_b/N_0$ required to achieve a target BER of $10^{-4}$ for a $\mathcal{P}$(2048,1024)-code over AWGN channel}\label{tab:decoder-tailored}
		\begin{tabular}{cccc}
			\hline 
			\multirow{2}{*}{\scriptsize Construction @ design SNR} & \multicolumn{3}{c}{\scriptsize Decoder}\tabularnewline
			\cline{2-4} 
			&\scriptsize SC & \scriptsize BP $\left(N_{it,max}=200\right)$ & \scriptsize SCL $\left(L=32\right)$ \tabularnewline
			\hline 
			\scriptsize Bhattacharyya \cite{ArikanMain} @ $\unit[3.6]{dB}$ & \scriptsize \textbf{2.7 dB} & \scriptsize 2.45 dB & \scriptsize 1.8 dB \tabularnewline
			\scriptsize Tal and Vardy \cite{constructTalVardy} @ $\unit[2]{dB}$ & \scriptsize \textbf{2.65 dB} & \scriptsize 2.45 dB & \scriptsize 2 dB \tabularnewline
			\scriptsize GenAlg BP-tailored @ $\unit[2]{dB}$ & \scriptsize $>$ 9 dB & \scriptsize \textbf{2 dB} & \scriptsize $>$ 7 dB\tabularnewline
			\scriptsize GenAlg SCL-tailored @ $\unit[2]{dB}$ & \scriptsize $>$ 6 dB & \scriptsize 2.55 dB & \scriptsize \textbf{1.65 dB}\tabularnewline
			\hline 
		\end{tabular}
	\end{center}
\end{table}

	\vspace{-0.5cm}	

\subsection{Weight enumerators $A_d$}

The weight enumerator $A_d$ is the number of codewords with weight $d$.
There is no closed-form expression for the weight enumerators of polar codes \cite{BP_LLR_Siegel}.
However, an algorithm proposed in \cite{AdaptiveList} can be used to find the number of minimum-weight codewords (i.e., $A_{d_{min}}$).
Table \ref{tab:A_dmin} shows the number of low-weight codewords in polar codes constructed with different methods. 
Comparing the number of minimum-weight codewords provides one explanation for the improved error-rate performance of the GenAlg-based polar codes. It can be seen that the genetic algorithm significantly reduces the number of minimum-weight codewords and the total number of low-weight codewords, as shown in Table \ref{tab:A_dmin}.

\begin{table}[h]
	\begin{center}
		\caption{The number of low-weight codewords for a $\mathcal{P}$(2048,1024)-code over AWGN channel}
		\label{tab:A_dmin}
		\begin{tabular}{cccc}
			\hline 
			Construction @ design SNR & $d_{min}$ & $A_8$ & $A_{16}$\tabularnewline
			\hline 
			Tal and Vardy \cite{constructTalVardy} @ $\unit[2]{dB}$ & 16 & 0 & 11648\tabularnewline
			GenAlg BP-tailored @ $\unit[2]{dB}$ & 8 & 8 & 773\tabularnewline
			GenAlg SCL-tailored @ $\unit[2]{dB}$ & 16 & 0 & 1\tabularnewline
			\hline 
		\end{tabular}
	\end{center}
\end{table}

Fig. \ref{fig:SCL_comp} shows curves with the same slope for different polar codes with different $d_{min}$. This observation can be attributed to the fact that the SCL ($L=32$) decoder is a sub-optimal decoder for RM-Polar codes \cite{HybridTse} and CRC-aided polar codes \cite{AdaptiveList} (i.e., a larger list size is needed to exploit the benefits of the enhanced $d_{min}$). Furthermore, the GenAlg SCL-tailored polar code has $d_{min}=16$ with only one codeword with weight $16$ (i.e., $A_{16}=1$). Thus, most of the decoding block errors are attributed to the weight-32 codewords, if we consider the case of transmitting the all-zero codeword.
	
\section{Results for the Rayleigh fading channel} \label{sec:resRayleigh}

The problem of designing or constructing polar codes for the Rayleigh fading channel was tackled in \cite{RayleighAngelBravo,TrifonovRayleigh}. To demonstrate the flexibility of the \ac{GenAlg}-based design, we optimize polar codes for Rayleigh fading channels and outline some performance comparisons. To be coherent with the results shown in \cite{TrifonovRayleigh}, we use codes of length $N=1024$ and code rate $R_c=0.5$ in the Rayleigh fading channel simulations. Following the notation in \cite{StBEXIT}, the system model is given by 

$$y=\alpha \cdot x + n$$

where $y$ is the channel output, $n$ is the Gaussian noise attributed to the channel such that $n\sim\mathcal{N}(0,\sigma_{ch}^2)$ and $\alpha > 0$ is the fading coefficient which follows a Rayleigh distribution with $E[\alpha^2]=1$.
Assuming perfect \ac{CSI} (i.e., $\alpha$ is known to the  receiver at each received bit position), the corresponding channel \acp{LLR}, namely $L_{ch}$, are computed as 
$$L_{ch}(y)=\log\frac{P\left(y|\alpha,x=+1\right)}{P\left(y|\alpha,x=-1\right)}=\frac{2}{\sigma_{ch}^2}\cdot \alpha \cdot y.$$
	
\subsection{\ac{BP} decoder}

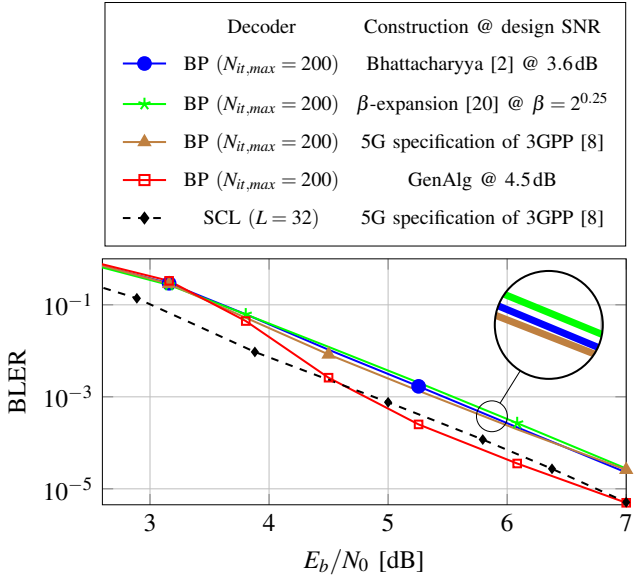
\begin{figure}[t]
	\centering
	\resizebox{0.975\columnwidth}{!}{\begin{tikzpicture}[spy using outlines={circle, magnification=3.5, connect spies}]
\begin{axis}[
width=\linewidth,
height=5cm,
xmajorgrids,
yminorticks=false,
ymajorgrids,
legend columns=1,
xlabel={$E_b/N_0$ [dB]},
ylabel={BLER},
ymode=log,
mark size=1.5pt,
xmin=2.6,
xmax=7,
ytick={1e-1, 1e-3, 1e-5},
ymin=4.5e-06,
ymax=1,
legend image post style={mark indices={}},
mark options={solid}
]

\addplot [color=blue,solid, mark=*, thick,mark indices={0,1,4},mark size=2.5pt] 
table[row sep=crcr]{
2	0.189486994219653\\
2.56078449083114	0.729820770306343\\
3.160288986114	0.294179197023717\\
3.80425961686212	0.0623246355304561\\
4.49982487708213	0.0105642411760912\\
5.25597799722985	0.00168895608714173\\
6.08429000539194	0.000216937713265591\\
7	2.26946062485816e-05\\
};
\label{plot:BP_Rayleigh_Bha}

\addplot [color=green,solid, mark=star, thick,mark indices={1,2,5},mark size=2.5pt] 
table[row sep=crcr]{
	2.00000000000000	0.956730439552904\\
	2.56078449083114	0.690575412726398\\
	3.16028898611400	0.270168230525790\\
	3.80425961686212	0.0620089127732003\\
	4.49982487708213	0.0121827807099472\\
	5.25597799722985	0.00202993910182695\\
	6.08429000539194	0.000268304737696816\\
	7	                2.75933476380751e-05\\
};
\label{plot:BP_Rayleigh_Beta}

\addplot [color=brown,solid, mark=triangle*, thick,mark indices={1,3,6},mark size=2.5pt] 
table[row sep=crcr]{
	2.00000000000000	0.971428571428571\\
	2.56078449083114	0.750494446980070\\
	3.16028898611400	0.285638220630234\\
	3.80425961686212	0.0524457001589264\\
	4.49982487708213	0.00833230288260114\\
	5.25597799722985	0.00135219986478001\\
	6.08429000539194	0.000196099588190865\\
	7	                2.59135864933985e-05\\ 
};
\label{plot:BP_Rayleigh_5G}

\addplot [color=red,solid, mark=square, thick] 
table[row sep=crcr]{
2.00000000000000	0.982162045991833\\
2.56078449083114	0.807266210492017\\
3.16028898611400	0.332238043081417\\
3.80425961686212	0.0444565771274660\\
4.49982487708213	0.00260564537129522\\
5.25597799722985	0.000251353039019557\\
6.08429000539194	3.55300937994476e-05\\
7	                4.97029091112703e-06\\
};
\label{plot:GenAlg_BP_Rayleigh}

\addplot [color=black,dashed, mark=diamond*, thick] 
table[row sep=crcr]{
	2.00000000000000	0.718997734994338\\
	2.88962579205211	0.138618980548771\\
	3.88087633102702	0.00945488366100439\\
	5	                0.000760398859401711\\
	5.79507557812516	0.000117797901836216\\
	6.37666082236955	2.72159049748673e-05\\ 
	7	                5.10227714629039e-06\\ 
};
\label{plot:SCL_Rayleigh_5G_2}

    \coordinate (legend) at (axis description cs:1,1.05);
\end{axis}

    \matrix [
    draw,
    matrix of nodes,
    anchor=south east,
    font=\footnotesize,
    mark options={solid}
    ] at (legend) {
    	& Decoder & Construction @ design SNR \\
    	\ref{plot:BP_Rayleigh_Bha} & BP $\left(N_{it,max}=200\right)$   & Bhattacharyya \cite{ArikanMain} @ $\unit[3.6]{dB}$   \\
    	\ref{plot:BP_Rayleigh_Beta} & BP $\left(N_{it,max}=200\right)$   & $\beta$-expansion \cite{BetaIngmard} @ $\beta=2^{0.25}$   \\
    	\ref{plot:BP_Rayleigh_5G} & BP $\left(N_{it,max}=200\right)$   & 5G specification of 3GPP \cite{polar5G2018}   \\    	    	
    	\ref{plot:GenAlg_BP_Rayleigh} & BP $\left(N_{it,max}=200\right)$   & GenAlg @ $\unit[4.5]{dB}$   \\
    	\ref{plot:SCL_Rayleigh_5G_2} & SCL $\left(L=32\right)$   & 5G specification of 3GPP \cite{polar5G2018}   \\  	
    };
\begin{scope}
\spy[black,size=1.5cm] on (5.41,1.23) in node [fill=none] at (6.2,2.5);
\end{scope}
\end{tikzpicture}}
	\caption{\small BLER performance of the \ac{GenAlg}-based $\mathcal{P}$(1024,512)-code under \ac{BP} decoding over the Rayleigh fading channel and no CRC is used.
		}		
	\label{fig:BP_Ray_comp}
			\vspace{-0.5cm}	
\end{figure}
	
We now design polar codes tailored to \ac{BP} decoding over the Rayleigh fading channel. Fig. \ref{fig:BP_Ray_comp} shows a \ac{BLER} comparison between a code constructed via \cite{BetaIngmard}, a code constructed via Arıkan's Bhattacharyya bounds \cite{ArikanMain} at $\unit[3.6]{dB}$, a code constructed based on the 5G specification of 3GPP \cite{polar5G2018} and a code constructed using our proposed \ac{GenAlg} at $SNR_{GenAlg}$ $\left({E_b}/{N_0}\right) = \unit[4.5]{dB}$ under \ac{BP} decoding with $N_{it,max} = 200$ iterations and the $\mathbf{G}$-matrix-based early stopping condition. The \ac{GenAlg}-based construction yields a $\unit[0.8]{dB}$ net coding gain at BLER of $10^{-4}$ when compared to the constructions proposed in \cite{ArikanMain,polar5G2018,BetaIngmard}. The above mentioned four $\mathbf{A}$-vectors yield different polar codes with $d_{min}=16$. It is remarkable to observe that the BLER performance of the \ac{BP}-tailored polar code (\ref{plot:GenAlg_BP_Rayleigh}) outperforms the performance of the proposed 5G polar code \cite{polar5G2018} under \ac{SCL} $\left(L=32\right)$ (\ref{plot:SCL_Rayleigh_5G_2}) starting from $\unit[4.5]{dB}$, as shown in Fig. \ref{fig:BP_Ray_comp}.
	
\subsection{\ac{SCL} decoder}

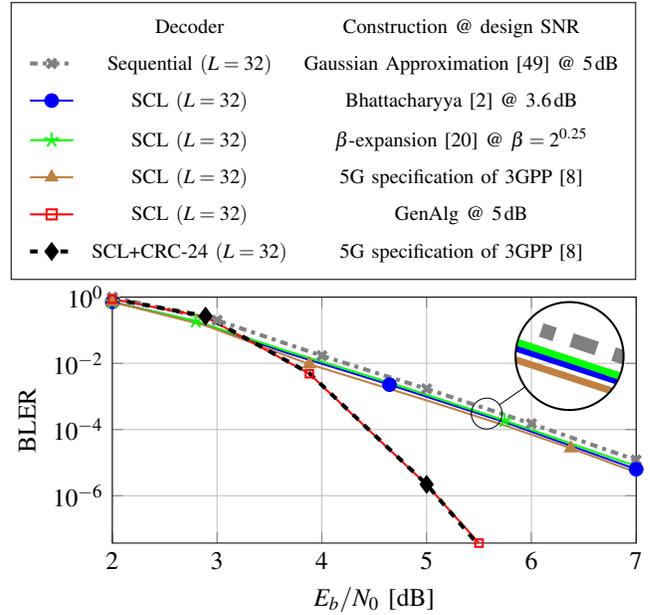
\begin{figure}[t]
	\centering
	\resizebox{0.975\columnwidth}{!}{\begin{tikzpicture}[spy using outlines={circle, magnification=3.5, connect spies}]
\begin{axis}[
width=\linewidth,
height=5cm,
xmajorgrids,
yminorticks=false,
ymajorgrids,
legend columns=1,
xlabel={$E_b/N_0$ [dB]},
ylabel={BLER},
ymode=log,
mark size=1.5pt,
xmin=2,
xmax=7,
ytick={1,1e-2, 1e-4, 1e-6,1e-8},
ymin=3.70003165007074e-08,
ymax=1,
legend image post style={mark indices={}},
mark options={solid}
]

\addplot [color=gray,solid, mark=x, ultra thick, dash dot,mark size=2.5pt] 
table[row sep=crcr]{
	2	1\\
	3	0.2\\
	4	0.017\\
	5	0.0017\\
	6	0.00015\\
	7	1.2e-05\\
};
\label{plot:Seq_Trifonov}

\addplot [color=blue,solid, mark=*, thick,mark indices={0,1,4,6,7,8,9,10},mark size=2.5pt] 
table[row sep=crcr]{
2	0.721371309391905\\
2.79564462090408	0.18480539699014\\
3.6715892883487	0.0204021016327449\\
4.64588085277352	0.0022707956854882\\
5.74342931819689	0.000167755476736157\\
7	6.31689384597216e-06\\
};
\label{plot:SCL_Rayleigh_Bha}

\addplot [color=green,solid, mark=star, thick,mark indices={1,2,5},mark size=3pt]
table[row sep=crcr]{
2	0.707628173220508\\
2.79564462090408	0.18873165567545\\
3.6715892883487	0.0231966782015664\\
4.64588085277352	0.0026306944755416\\
5.74342931819689	0.000188336826704601\\
7	7.8026583657052e-06\\
};
\label{plot:SCL_Rayleigh_beta}

\addplot [color=brown,solid, mark=triangle*, thick,mark indices={1,3,7},mark size=2.5pt] 
table[row sep=crcr]{
	2.00000000000000	0.718997734994338\\
	2.88962579205211	0.138618980548771\\
	3.88087633102702	0.00945488366100439\\
	5	                0.000760398859401711\\
	5.25000000000000	0.000427622236356291\\
	5.79507557812516	0.000117797901836216\\
	6.37666082236955	2.72159049748673e-05\\ 
	7	                5.10227714629039e-06\\ 
};
\label{plot:SCL_Rayleigh_5G}

\addplot [color=red,solid, mark=square, thick,mark indices={1,3,5}] 
table[row sep=crcr]{
	2.00000000000000	0.874201664408748\\
	2.88962579205211	0.253524403294096\\
	3.88087633102702	0.00486121052111173\\
	5	                2.14402133944439e-06\\ 
	5.50000000000000	3.70003165007074e-08\\ 
};
\label{plot:SCL_Rayleigh_GenAlg}

\addplot [color=black,solid, mark=diamond*, ultra thick, dashed,mark indices={2,4,6},mark size=2.5pt] 
table[row sep=crcr]{
	2.00000000000000	0.884393063583815\\
	2.88962579205211	0.266924969893712\\
	3.88087633102702	0.00502587583987804\\
	5	                2.20070884832004e-06\\ 
	5.50000000000000	3.07617369188820e-08\\ 
};
\label{plot:SCL_CRC_5G_Ray} 

    \coordinate (legend) at (axis description cs:1,1.07);
\end{axis}

    \matrix [
    draw,
    matrix of nodes,
    anchor=south east,
    font=\footnotesize,
    mark options={solid}
    ] at (legend) {
    	& Decoder & Construction @ design SNR \\
    	\ref{plot:Seq_Trifonov} & Sequential $\left(L=32\right)$   & Gaussian Approximation \cite{TrifonovRayleigh} @ $\unit[5]{dB}$   \\
    	\ref{plot:SCL_Rayleigh_Bha} & SCL $\left(L=32\right)$   & Bhattacharyya \cite{ArikanMain} @ $\unit[3.6]{dB}$   \\
    	\ref{plot:SCL_Rayleigh_beta} & SCL $\left(L=32\right)$   & $\beta$-expansion \cite{BetaIngmard} @ $\beta=2^{0.25}$   \\
    	\ref{plot:SCL_Rayleigh_5G} & SCL $\left(L=32\right)$   & 5G specification of 3GPP \cite{polar5G2018}   \\  	
    	\ref{plot:SCL_Rayleigh_GenAlg} & SCL $\left(L=32\right)$   & GenAlg @ $\unit[5]{dB}$   \\    	
 		\ref{plot:SCL_CRC_5G_Ray} & SCL+CRC-24 $\left(L=32\right)$   &  5G specification of 3GPP \cite{polar5G2018}  \\		   	
    };
\begin{scope}
\spy[black,size=1.5cm] on (5.2,1.8) in node [fill=none] at (6.35,2.6);
\end{scope}
\end{tikzpicture}}
	\caption{\small BLER performance of the \ac{GenAlg}-based $\mathcal{P}$(1024,512)-code under \ac{SCL} decoding over the Rayleigh fading channel. The CRC-aided polar code (\ref{plot:SCL_CRC_5G_Ray}): $N=1024$, $k=512$, $r=24$ and, thus, the polar code is a $\mathcal{P}$(1024,536)-code.}		
	\label{fig:SCL_Ray_comp}
			\vspace{-0.5cm}	
\end{figure}
	
In this section, polar codes are tailored to \ac{SCL} $\left(L=32\right)$ decoding over the Rayleigh fading channel using the \ac{GenAlg}. Fig. \ref{fig:SCL_Ray_comp} shows a \ac{BLER} comparison between a code constructed via \cite{BetaIngmard}, a code constructed via Arıkan's Bhattacharyya bounds \cite{ArikanMain} at $\unit[3.6]{dB}$, a code constructed based on the 5G specification of 3GPP \cite{polar5G2018} and a code constructed using our proposed \ac{GenAlg} at $SNR_{GenAlg}$ $\left({E_b}/{N_0}\right) = \unit[5]{dB}$ under \ac{SCL} $\left(L=32\right)$ decoding. \ac{GenAlg}-based construction yields a $\unit[2]{dB}$ net coding gain at BLER of $10^{-6}$ when compared to the constructions proposed in \cite{ArikanMain,polar5G2018,BetaIngmard}. It is worth mentioning that our proposed SCL-tailored polar code (\ref{plot:SCL_Rayleigh_GenAlg}) performs equally well, in terms of error-rate, as the best polar code with dynamic frozen symbols in \cite[Fig. 4]{TrifonovRayleigh} for the same list size $L=32$. Also, our proposed code without the aid of a CRC performs equally well, in terms of error-rate, as the proposed 5G polar code \cite{polar5G2018} under CRC-aided SCL decoding (\ref{plot:SCL_CRC_5G_Ray}). The GenAlg constructed polar code tailored to SCL has a $d_{min}=32$.

Note that the above shown gains come ``for free'', since optimizing the $\mathbf{A}$-vector is done offline and it does not affect the decoding complexity, latency or memory requirements. However, for a fixed target error-rate, the decoding complexity, latency and memory requirements can also be reduced using \ac{GenAlg} as shown in the next section.

\section{Decoding Complexity Reduction} \label{sec:complexity}

Besides error-rate performance, also the decoding complexity, memory requirements and latency are of major significance when selecting the \emph{best} channel coding scheme for a certain application. Further investigations are conducted to study the impact of the polar code construction using the \ac{GenAlg} on the decoding complexity. This problem can be formulated as follows. We try to construct a polar code using the \ac{GenAlg} such that, for a given fixed target error-rate level, the decoder used to decode the GenAlg-designed polar code is significantly less complex than the conventional polar code decoder.
This complexity reduction is of potential importance when it comes to practical applications with constraints on the decoding complexity, latency and memory requirements.
	
\subsection{\ac{BP} decoder} 

\begin{figure}[t]
	\centering
	\resizebox{0.975\columnwidth}{!}{\begin{tikzpicture}
\begin{axis}[
width=\linewidth,
height=5cm,
xmajorgrids,
yminorticks=false,
ymajorgrids,
legend columns=1,
xlabel={$E_b/N_0$ [dB]},
ylabel={BER},
ymode=log,
mark size=1.5pt,
xmin=1.5,
xmax=3,
xtick={1.5,1.8,2.1,2.4,2.7,3},
ytick={1e-1,1e-2, 1e-3, 1e-4, 1e-5},
ymin=1e-5,
ymax=4e-2,
]

\addplot [color=blue,solid, mark=diamond*, thick] 
table[row sep=crcr]{
	1.00000000000000	0.112095924497848\\
	1.24479414452514	0.0371581009546795\\
	1.49668788567703	0.0105970127808852\\
	1.75610536460180	0.00282372360618388\\
	2.02350990413076	0.000753127703832328\\
	2.29940898861021	0.000203934403722672\\
	2.58436006091173	5.73887868431984e-05\\
	2.87897730301695	1.64008483608058e-05\\
	3.18393960749088	4.96658765605456e-06\\
	3.50000000000000	1.23719593571999e-06\\
};
\label{plot:BP_TalVardy_Comp_red}

\addplot [color=green,solid, mark=*, thick] 
table[row sep=crcr]{
	1.00000000000000	0.184313651842949\\
	1.24479414452514	0.0829902220118739\\
	1.49668788567703	0.0288233936450432\\
	1.75610536460180	0.00798220422260022\\
	2.02350990413076	0.00205278290436985\\
	2.29940898861021	0.000484786788956820\\
	2.58436006091173	0.000125068647546414\\
	2.87897730301695	3.16939072798113e-05\\
	3.18393960749088	7.35347194460837e-06\\
	3.50000000000000	1.94604617641223e-06\\
};
\label{plot:BP_TalVardy_30}

\addplot [color=red,solid, mark=star, thick] 
table[row sep=crcr]{
	1.00000000000000	0.170932265242302\\
	1.24479414452514	0.0743405694640003\\
	1.49668788567703	0.0217115582867432\\
	1.75610536460180	0.00438736227138580\\
	2.02350990413076	0.000757534209883318\\
	2.29940898861021	0.000149926555615639\\
	2.58436006091173	4.17526759212537e-05\\
	2.87897730301695	1.41767365473952e-05\\
	3.18393960749088	5.34373236568319e-06\\
	3.50000000000000	2.20312147500564e-06\\
};
\label{BP_GenAlg_30}

    \coordinate (legend) at (axis description cs:0.985,1.05);
\end{axis}

    \matrix [
    draw,
    matrix of nodes,
    anchor=south east,
    font=\footnotesize,
    ] at (legend) {
    	& Decoder & Construction @ design SNR \\
    	\ref{plot:BP_TalVardy_Comp_red} & BP $\left(N_{it,max}=200\right)$   & Tal and Vardy \cite{constructTalVardy} @ $\unit[2]{dB}$   \\
    	\ref{plot:BP_TalVardy_30} & BP $\left(N_{it,max}=30\right)$   & Tal and Vardy \cite{constructTalVardy} @ $\unit[2]{dB}$   \\    	
    	\ref{BP_GenAlg_30} & BP $\left(N_{it,max}=30\right)$   & GenAlg @ $\unit[2]{dB}$   \\
    };

\end{tikzpicture}}
	\caption{\small BER performance of the \ac{GenAlg}-based $\mathcal{P}$(2048,1024)-code under \ac{BP} decoding with reduced $N_{it,max}$ over the \ac{AWGN} channel and no CRC is used.
		}		
	\label{fig:BP_comp_red}
		\vspace{-0.5cm}	
\end{figure}
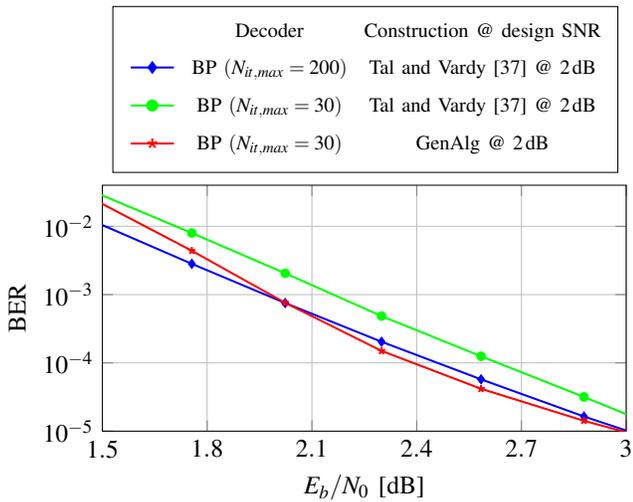

The \ac{GenAlg}-based \ac{BP}-tailored polar code only requires $30$ \ac{BP} iterations to achieve the same error-rate performance as the polar code constructed via \cite{constructTalVardy} under $200$ \ac{BP} iterations, as shown in Fig. \ref{fig:BP_comp_red}. Thus, an $85 \%$ reduction in the worst case decoding complexity and latency (remember the early stopping condition used) can be achieved with the same error-rate performance just by changing the set of frozen/non-frozen bit positions. Furthermore, at $E_b/N_0=\unit[2]{dB}$ a $5\%$ reduction in the average number of BP iterations is achieved with the same error-rate performance, reducing the average decoding complexity and latency by $5\%$.
	
\subsection{\ac{SCL} decoder} \label{sssec:SCLcomplexity}
The \ac{GenAlg}-based \ac{SCL}-tailored polar code only requires a list size of $4$ to outperform, in terms of error-rate, the polar code constructed via \cite{constructTalVardy} under \ac{SCL} with list size $32$, as shown in Fig. \ref{fig:SCL_comp_red}. Thus, an $87.5 \%$ reduction in the decoding complexity and memory requirements with an improved error-rate performance can be achieved just by tailoring the set of frozen/non-frozen bit positions to the decoder.
Note that a larger list size tends to cause higher decoding latency \cite{hardwarePolarList,fastListDecoder} and thus, the optimized polar code construction also reduces the decoding latency.
It should be emphasized that no additional CRC was used for the results in Fig. \ref{fig:SCL_comp_red}.

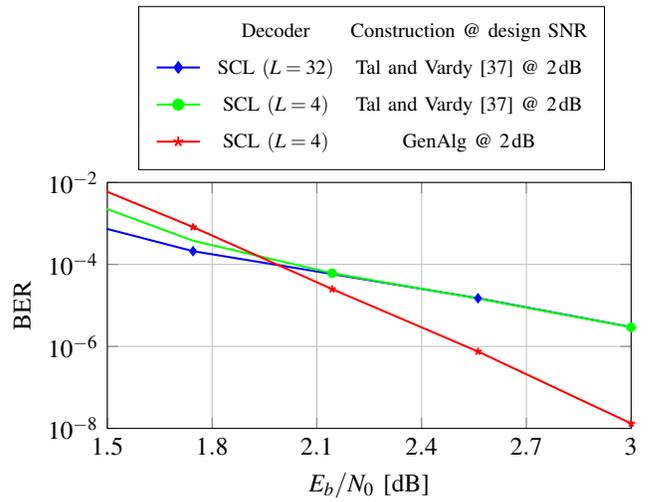
\begin{figure}[t]
	\centering
	\resizebox{0.975\columnwidth}{!}{\begin{tikzpicture}
\begin{axis}[
width=\linewidth,
height=5cm,
xmajorgrids,
yminorticks=false,
ymajorgrids,
xlabel={$E_b/N_0$ [dB]},
ylabel={BER},
ymode=log,
mark size=1.5pt,
xmin=1.5,
xmax=3,
xtick={1.5,1.8,2.1,2.4,2.7,3},
ytick={1e-2, 1e-4, 1e-6, 1e-8},
ymin=1e-8,
ymax=1e-2,
]

  \addplot [color=blue,solid, mark=diamond*, thick,mark repeat=2,mark phase=0] 
table[row sep=crcr]{
	1.00000000000000	0.0209942992156605\\
	1.36484480920620	0.00145650969990947\\
	1.74568917910050	0.000210496604956101\\
	2.14400098577864	5.84599609375000e-05\\
	2.56145986298185	1.49636509258887e-05\\
	3.00000000000000	2.95790542360132e-06\\
};
\label{plot:SCL_TalVardy_Comp_red}

  \addplot [color=green,solid, mark=*, mark options={scale=1}, thick,mark repeat=2,mark phase=2] 
  table[row sep=crcr]{
1.00000000000000	0.0636708945858731\\
1.36484480920620	0.00597353317496774\\
1.74568917910050	0.000378305693402379\\
2.14400098577864	6.11882259356852e-05\\
2.56145986298185	1.46670634430358e-05\\
3.00000000000000	2.92411407587622e-06\\
};
\label{plot:SCL_TalVardy_L4}

  \addplot [color=red,solid, mark=star, thick] 
  table[row sep=crcr]{
1.00000000000000	0.124063435138589\\
1.36484480920620	0.0176311039188500\\
1.74568917910050	0.000815183482926662\\
2.14400098577864	2.49800306629417e-05\\
2.56145986298185	7.51231632675027e-07\\
3.00000000000000	1.31751296960627e-08\\
  };
  \label{plot:GenAlg_SCL_4}

\coordinate (legend) at (axis description cs:0.95,1.05);
\end{axis}

    \matrix [
    draw,
    matrix of nodes,
    anchor=south east,
    font=\footnotesize,
    ] at (legend) {
    	& Decoder & Construction @ design SNR \\
    	\ref{plot:SCL_TalVardy_Comp_red} & SCL $\left(L=32\right)$   & Tal and Vardy \cite{constructTalVardy} @ $\unit[2]{dB}$   \\    	
    	\ref{plot:SCL_TalVardy_L4} & SCL $\left(L=4\right)$   & Tal and Vardy \cite{constructTalVardy} @ $\unit[2]{dB}$   \\
    	\ref{plot:GenAlg_SCL_4} & SCL $\left(L=4\right)$   & GenAlg @ $\unit[2]{dB}$   \\    	
    };

\end{tikzpicture}}
	\caption{\small BER performance of the \ac{GenAlg}-based $\mathcal{P}$(2048,1024)-code under \ac{SCL} decoding with reduced list size $L$ over the \ac{AWGN} channel and no CRC is used.
		}		
	\label{fig:SCL_comp_red}
	\vspace{-0.5cm}
\end{figure}

Despite the \emph{offline} construction complexity added by the \ac{GenAlg}, the resulting polar codes can be decoded with state-of-the-art decoders with much lower decoding complexity, latency and memory requirements at a fixed target error-rate, when compared to polar codes constructed via other state-of-the-art polar code construction techniques (e.g., \cite{constructTalVardy}). Thus, it is a trade-off between offline complexity (i.e., polar code construction complexity) and online complexity (i.e., polar decoding complexity).

According to \cite{Marco_Sublinear}, $80\%$ of the channel (i.e., bit-channel reliability) computations can be saved for moderate block length polar codes (e.g., $N \approx 1000$). 
Thus, we believe that the GenAlg-based polar code construction complexity can be significantly reduced by fixing some good bit-channels as “for sure non-frozen” and some bad bit-channels as “for sure frozen”, leading to obvious speed-ups in the convergence speed of the GenAlg.
The needed prior knowledge can be easily derived from \cite{UPO,Marco_Sublinear,BetaIngmard}.

\section{Conclusion} \label{sec:conc}
As pointed out in \cite{talvardyList}, the best error-rate performance of finite length polar codes (i.e., under \ac{ML} decoding without \ac{CRC} concatenation) is not competitive with the state-of-the-art coding schemes. Thus in this work, we focus on polar code construction (i.e., changing the code itself which is defined by the $\mathbf{A}$-vector) in order to boost the error-rate performance under the state-of-the-art feasible practical decoders (e.g., \ac{BP}, \ac{SCL}). We propose a Genetic Algorithm-based polar code design where the channel type and the nature of the decoder are taken into consideration, yielding significant performance gains in terms of error-rate. This also shows the importance of a proper design algorithm and provides an estimate of the capabilities of well-suited polar codes.
For a polar code of length $2048$ and code rate $0.5$ over the binary input \ac{AWGN} channel under \emph{plain} \ac{SCL} decoding, approximately a $\unit[1]{dB}$ coding gain at \ac{BER} of $10^{-6}$ is achieved when compared to the conventionally constructed polar codes. This enables achieving the same error-rate performance of polar codes under state-of-the-art CRC-aided SCL decoding with \emph{plain} \ac{SCL} decoding without the aid of a CRC. 
Furthermore, \ac{GenAlg} is used to construct polar codes tailored to flooding \ac{BP} decoding, finally closing the performance gap between conventional iterative \ac{BP} decoding and conventional \ac{SCL} decoding.

\ac{GenAlg} was used to reduce the decoding complexity, latency and memory requirements of the \ac{BP} and \ac{SCL} decoders. Thus, after \ac{GenAlg}-based $\mathbf{A}$-vector optimization, a few number of \ac{BP} iterations or a small list size is needed, in order to achieve a fixed target error-rate.

In addition, for a polar code of length $1024$ and code rate $0.5$, \ac{GenAlg} is used to optimize polar codes over Rayleigh fading channels, yielding a $\unit[2]{dB}$ performance gain under \ac{SCL} decoding at \ac{BLER} of $10^{-6}$ when compared to conventional polar design methods. The results suggest that the improved performance is due to the significant reduction in the total number of low-weight codewords. It is worth mentioning that the achieved gains come ``for free'', since optimizing the $\mathbf{A}$-vector is done once and offline.

\section{Acknowledgment}
The authors would like to thank Alexander Vardy and Ido Tal for providing their set of frozen/non-frozen bit positions as a baseline for comparison.
	
\bibliographystyle{IEEEtran}
\bibliography{references}

\begin{thebibliography}{10}
\providecommand{\url}[1]{#1}
\csname url@samestyle\endcsname
\providecommand{\newblock}{\relax}
\providecommand{\bibinfo}[2]{#2}
\providecommand{\BIBentrySTDinterwordspacing}{\spaceskip=0pt\relax}
\providecommand{\BIBentryALTinterwordstretchfactor}{4}
\providecommand{\BIBentryALTinterwordspacing}{\spaceskip=\fontdimen2\font plus
\BIBentryALTinterwordstretchfactor\fontdimen3\font minus
  \fontdimen4\font\relax}
\providecommand{\BIBforeignlanguage}[2]{{%
\expandafter\ifx\csname l@#1\endcsname\relax
\typeout{** WARNING: IEEEtran.bst: No hyphenation pattern has been}%
\typeout{** loaded for the language `#1'. Using the pattern for}%
\typeout{** the default language instead.}%
\else
\language=\csname l@#1\endcsname
\fi
#2}}
\providecommand{\BIBdecl}{\relax}
\BIBdecl

\bibitem{GenAlgSCC19}
A.~Elkelesh, M.~Ebada, S.~Cammerer, and S.~ten Brink, ``{Genetic
  Algorithm-based Polar Code Construction for the AWGN Channel},'' in
  \emph{IEEE Inter. ITG Conf. on Syst., Commun. and Coding (SCC)}, Feb. 2019.

\bibitem{ArikanMain}
E.~{Arıkan}, ``{Channel Polarization: A Method for Constructing
  Capacity-Achieving Codes for Symmetric Binary-Input Memoryless Channels},''
  \emph{IEEE Trans. Inf. Theory}, vol.~55, no.~7, pp. 3051--3073, July 2009.

\bibitem{polar_comparisons}
A.~Balatsoukas-Stimming, P.~Giard, and A.~Burg, ``{Comparison of Polar Decoders
  with Existing Low-Density Parity-Check and Turbo Decoders},'' in \emph{IEEE
  Wireless Commun. and Networking Conf. (WCNC)}, Mar. 2017.

\bibitem{talvardyList}
I.~Tal and A.~Vardy, ``{List Decoding of Polar Codes},'' \emph{IEEE Trans. Inf.
  Theory}, vol.~61, no.~5, pp. 2213--2226, May 2015.

\bibitem{miloslavskaya2014sequential}
V.~Miloslavskaya and P.~Trifonov, ``{Sequential Decoding of Polar Codes},''
  \emph{IEEE Commun. Lett.}, vol.~18, no.~7, pp. 1127--1130, July 2014.

\bibitem{PCC}
T.~Wang, D.~Qu, and T.~Jiang, ``{Parity-Check-Concatenated Polar Codes},''
  \emph{IEEE Commun. Lett.}, vol.~20, no.~12, pp. 2342--2345, Dec. 2016.

\bibitem{liva2016}
G.~Liva, L.~Gaudio, T.~Ninacs, and T.~Jerkovits, ``{Code Design for Short
  Blocks: A Survey},'' \emph{ArXiv e-prints}, Oct. 2016.

\bibitem{polar5G2018}
\BIBentryALTinterwordspacing
``{Technical Specification Group Radio Access Network},'' \emph{3GPP, 2018, TS
  38.212 V.15.1.1.} [Online]. Available:
  \url{http://www.3gpp.org/ftp/Specs/archive/38_series/38.212/}
\BIBentrySTDinterwordspacing

\bibitem{HW}
P.~Giard, G.~Sarkis, A.~Balatsoukas-Stimming, Y.~Fan, C.~y.~Tsui, A.~Burg,
  C.~Thibeault, and W.~J. Gross, ``{Hardware Decoders for Polar Codes: An
  Overview},'' in \emph{IEEE Inter. Symp. on Circuits Syst. (ISCAS)}, May 2016,
  pp. 149--152.

\bibitem{ArikanBP}
E.~Arıkan, ``{A Performance Comparison of Polar Codes and Reed-Muller
  Codes},'' \emph{IEEE Commun. Lett.}, vol.~12, no.~6, pp. 447--449, June 2008.

\bibitem{elkelesh2018belief}
A.~Elkelesh, M.~Ebada, S.~Cammerer, and S.~ten Brink, ``{Belief Propagation
  List Decoding of Polar Codes},'' \emph{IEEE Commun. Lett.}, vol.~22, no.~8,
  pp. 1536--1539, Aug. 2018.

\bibitem{polarCodes_concepts}
K.~Niu, K.~Chen, J.~Lin, and Q.~T. Zhang, ``{Polar Codes: Primary Concepts and
  Practical Decoding Algorithms},'' \emph{IEEE Commun. Mag.}, vol.~52, no.~7,
  pp. 192--203, July 2014.

\bibitem{PSCL_Journal}
S.~A. Hashemi, M.~Mondelli, S.~H. Hassani, C.~Condo, R.~L. Urbanke, and W.~J.
  Gross, ``{Decoder Partitioning: Towards Practical List Decoding of Polar
  Codes},'' \emph{IEEE Trans. Commun.}, vol.~66, no.~9, pp. 3749--3759, Sep.
  2018.

\bibitem{ISWCS_Error_Floor}
A.~Elkelesh, S.~Cammerer, M.~Ebada, and S.~ten Brink, ``{Mitigating Clipping
  Effects on Error Floors under Belief Propagation Decoding of Polar Codes},''
  in \emph{Inter. Symp. Wireless Commun. Syst. (ISWCS)}, Aug. 2017.

\bibitem{constructDE}
R.~Mori and T.~Tanaka, ``{Performance of Polar Codes with the Construction
  using Density Evolution},'' \emph{IEEE Commun. Lett.}, vol.~13, no.~7, pp.
  519--521, July 2009.

\bibitem{constructGaussian}
P.~Trifonov, ``{Efficient Design and Decoding of Polar Codes},'' \emph{IEEE
  Trans. Commun.}, vol.~60, no.~11, pp. 3221--3227, Nov. 2012.

\bibitem{GA}
D.~Wu, Y.~Li, and Y.~Sun, ``{Construction and Block Error Rate Analysis of
  Polar Codes Over AWGN Channel Based on Gaussian Approximation},'' \emph{IEEE
  Commun. Lett.}, vol.~18, no.~7, pp. 1099--1102, July 2014.

\bibitem{PW}
``{3GPP, R1-167209, Polar code design and rate matching, Huawei, HiSilicon}.''

\bibitem{TUM_SCL_Construct}
P.~Yuan, T.~Prinz, and G.~Böcherer, ``{Polar Code Construction for List
  Decoding},'' \emph{ArXiv e-prints}, July 2017.

\bibitem{BetaIngmard}
G.~He, J.~C. Belfiore, I.~Land, G.~Yang, X.~Liu, Y.~Chen, R.~Li, J.~Wang,
  Y.~Ge, R.~Zhang, and W.~Tong, ``{$\beta$-expansion: A Theoretical Framework
  for Fast and Recursive Construction of Polar Codes},'' in \emph{IEEE Global
  Commun. Conf. (GLOBECOM)}, Dec. 2017, pp. 1--6.

\bibitem{MC_BP}
J.~Liu and J.~Sha, ``{Frozen bits selection for polar codes based on simulation
  and BP decoding},'' \emph{IEICE Electronics Express}, Mar. 2017.

\bibitem{MC_BP_2}
S.~Sun and Z.~Zhang, ``{Designing Practical Polar Codes Using Simulation-Based
  Bit Selection},'' \emph{IEEE J. Emerging and Sel. Topics Circuits Syst.},
  vol.~7, no.~4, pp. 594--603, Dec. 2017.

\bibitem{BP_LLR_Siegel}
M.~Qin, J.~Guo, A.~Bhatia, A.~G. i~Fabregas, and P.~Siegel, ``{Polar Code
  Constructions Based on LLR Evolution},'' \emph{IEEE Commun. Lett.}, vol.~21,
  no.~6, pp. 1221--1224, June 2017.

\bibitem{HybridTse}
B.~Li, H.~Shen, and D.~Tse, ``{A RM-Polar Codes},'' \emph{ArXiv e-prints}, July
  2014.

\bibitem{subCodes}
P.~Trifonov and V.~Miloslavskaya, ``{Polar Subcodes},'' \emph{IEEE J. Sel.
  Areas Commun.}, vol.~34, no.~2, pp. 254--266, Feb. 2016.

\bibitem{BP_felxible}
A.~Elkelesh, M.~Ebada, S.~Cammerer, and S.~ten Brink, ``{Flexible Length Polar
  Codes through Graph Based Augmentation},'' in \emph{IEEE Inter. ITG Conf. on
  Syst., Commun. and Coding (SCC)}, Feb. 2017.

\bibitem{BP_Siegel_Concatenating}
J.~Guo, M.~Qin, A.~G. i~Fàbregas, and P.~H. Siegel, ``{Enhanced Belief
  Propagation Decoding of Polar Codes through Concatenation},'' in \emph{IEEE
  Inter. Symp. Inf. Theory (ISIT)}, June 2014, pp. 2987--2991.

\bibitem{BP_sEXIT}
A.~Elkelesh, M.~Ebada, S.~Cammerer, and S.~ten Brink, ``{Improving Belief
  Propagation Decoding of Polar Codes Using Scattered EXIT Charts},'' in
  \emph{IEEE Inf. Theory Workshop (ITW)}, Sep. 2016, pp. 91--95.

\bibitem{arikan2011systematic}
E.~Arikan, ``{Systematic Polar Coding},'' \emph{IEEE Commun. Lett.}, vol.~15,
  no.~8, pp. 860--862, Aug. 2011.

\bibitem{earlyStop}
B.~Yuan and K.~K. Parhi, ``{Early Stopping Criteria for Energy-Efficient
  Low-Latency Belief-Propagation Polar Code Decoders},'' \emph{IEEE Trans. Sig.
  Process.}, vol.~62, no.~24, pp. 6496--6506, Dec. 2014.

\bibitem{UniversalPolarization}
E.~Şaşoğlu and L.~Wang, ``{Universal Polarization},'' \emph{IEEE Trans. Inf.
  Theory}, vol.~62, no.~6, pp. 2937--2946, June 2016.

\bibitem{Urbanke_chCsC_BP}
N.~Hussami, S.~B. Korada, and R.~Urbanke, ``{Performance of Polar Codes for
  Channel and Source Coding},'' in \emph{IEEE Inter. Symp. Inf. Theory (ISIT)},
  June 2009, pp. 1488--1492.

\bibitem{Vangala}
H.~Vangala, E.~Viterbo, and Y.~Hong, ``{A Comparative Study of Polar Code
  Constructions for the AWGN Channel},'' \emph{ArXiv e-prints}, Jan. 2015.

\bibitem{RMurbankePolar}
M.~Mondelli, S.~H. Hassani, and R.~L. Urbanke, ``{From Polar to Reed-Muller
  Codes: A Technique to Improve the Finite-Length Performance},'' \emph{IEEE
  Trans. Commun.}, vol.~62, no.~9, pp. 3084--3091, Sep. 2014.

\bibitem{polarDesign5G}
V.~Bioglio, C.~Condo, and I.~Land, ``{Design of Polar Codes in 5G New Radio},''
  \emph{ArXiv e-prints}, Apr. 2018.

\bibitem{minDisLand}
V.~Bioglio, F.~Gabry, I.~Land, and J.-C. Belfiore, ``{Minimum-Distance Based
  Construction of Multi-Kernel Polar Codes},'' in \emph{IEEE Global Commun.
  Conf. (GLOBECOM)}, Dec. 2017.

\bibitem{constructTalVardy}
I.~Tal and A.~Vardy, ``{How to Construct Polar Codes},'' \emph{IEEE Trans. Inf.
  Theory}, vol.~59, no.~10, pp. 6562--6582, Oct. 2013.

\bibitem{UPO}
C.~Schürch, ``{A Partial Order For the Synthesized Channels of a Polar
  Code},'' in \emph{IEEE Inter. Symp. Inf. Theory (ISIT)}, July 2016, pp.
  220--224.

\bibitem{Marco_Sublinear}
M.~Mondelli, S.~H. Hassani, and R.~Urbanke, ``{Construction of Polar Codes with
  Sublinear Complexity},'' \emph{IEEE Trans. Inf. Theory}, 2018.

\bibitem{GeneticsFirstpaper}
J.~H. Holland, \emph{Adaptation in Natural and Artificial Systems: An
  Introductory Analysis with Applications to Biology, Control and Artificial
  Intelligence}.\hskip 1em plus 0.5em minus 0.4em\relax Cambridge, MA, USA: MIT
  Press, 1992.

\bibitem{GAturbo}
L.~Hebbes, R.~R. Malyan, and A.~P. Lenaghan, ``Genetic algorithms for turbo
  codes,'' in \emph{EUROCON - The Inter. Conf. on Computer as a Tool}, vol.~1,
  Nov. 2005, pp. 478--481.

\bibitem{GAMRD}
K.~Dontas and K.~D. Jong, ``Discovery of maximal distance codes using genetic
  algorithms,'' in \emph{IEEE Inter. Conf. on Tools for Artificial
  Intelligence}, Nov. 1990, pp. 805--811.

\bibitem{GAlinearBlockCodes}
H.~Maini, K.~Mehrotra, C.~Mohan, and S.~Ranka, ``{Genetic Algorithms for
  Soft-decision Decoding of Linear Block Codes},'' \emph{Evol. Comput.},
  vol.~2, no.~2, pp. 145--164, June 1994.

\bibitem{GAldpcdec}
{A. Scandurra et al.}, ``{A Genetic-Algorithm Based Decoder for Low Density
  Parity Check Codes},'' \emph{Latin American Applied Research}, vol.~36,
  no.~3, pp. 169--172, July 2006.

\bibitem{GAconv}
H.~Berbia, M.~Belkasmi, F.~Elbouanani, and F.~Ayoub, ``On the decoding of
  convolutional codes using genetic algorithms,'' in \emph{Inter. Conf. on
  Computer and Commun. Engineering}, May 2008, pp. 667--671.

\bibitem{GAsurvey}
M.~Srinivas and L.~M. Patnaik, ``Genetic algorithms: a survey,''
  \emph{Computer}, vol.~27, no.~6, pp. 17--26, June 1994.

\bibitem{GAtruncation}
T.~Blickle and L.~Thiele, ``{A Comparison of Selection Schemes Used in
  Evolutionary Algorithms},'' \emph{Evolutionary Computation}, vol.~4, no.~4,
  pp. 361--394, Dec. 1996.

\bibitem{GApreConv}
L.~Juan, C.~Zixing, and L.~Jianqin, ``Premature convergence in genetic
  algorithm: analysis and prevention based on chaos operator,'' in \emph{Proc.
  3rd World Congress on Intelligent Control and Automation (Cat. No.00EX393)},
  vol.~1, June 2000, pp. 495--499 vol.1.

\bibitem{TrifonovRayleigh}
P.~Trifonov, ``{Design of Polar Codes for Rayleigh Fading Channel},'' in
  \emph{Inter. Symp. Wireless Commun. Syst. (ISWCS)}, Aug. 2015, pp. 331--335.

\bibitem{cammerer_HybridGPU}
S.~Cammerer, B.~Leible, M.~Stahl, J.~Hoydis, and S.~ten Brink, ``{Combining
  Belief Propagation and Successive Cancellation List Decoding of Polar Codes
  on a GPU Platform},'' in \emph{IEEE Inter. Conf. on Acoustics, Speech, and
  Sig. Process. (ICASSP)}, Mar. 2017, pp. 3664--3668.

\bibitem{vardyStoppingDistance}
M.~Schwartz and A.~Vardy, ``{On the Stopping Distance and the Stopping
  Redundancy of Codes},'' \emph{IEEE Trans. Inf. Theory}, vol.~52, no.~3, pp.
  922--932, Mar. 2006.

\bibitem{AdaptiveList}
B.~Li, H.~Shen, and D.~Tse, ``{An Adaptive Successive Cancellation List Decoder
  for Polar Codes with Cyclic Redundancy Check},'' \emph{IEEE Commun. Lett.},
  vol.~16, no.~12, pp. 2044--2047, Dec. 2012.

\bibitem{RayleighAngelBravo}
A.~Bravo-Santos, ``{Polar Codes for the Rayleigh Fading Channel},'' \emph{IEEE
  Commun. Lett.}, vol.~17, no.~12, pp. 2352--2355, Dec. 2013.

\bibitem{StBEXIT}
S.~ten Brink, ``{Convergence Behavior of Iteratively Decoded Parallel
  Concatenated Codes},'' \emph{IEEE Trans. Commun.}, vol.~49, no.~10, pp.
  1727--1737, Oct. 2001.

\bibitem{hardwarePolarList}
A.~Balatsoukas-Stimming, A.~J. Raymond, W.~J. Gross, and A.~Burg, ``{Hardware
  Architecture for List Successive Cancellation Decoding of Polar Codes},''
  \emph{IEEE Trans. Circuits Syst. II, Exp. Briefs}, vol.~61, no.~8, pp.
  609--613, Aug. 2014.

\bibitem{fastListDecoder}
G.~Sarkis, P.~Giard, A.~Vardy, C.~Thibeault, and W.~J. Gross, ``{Fast List
  Decoders for Polar Codes},'' \emph{IEEE J. Sel. Areas Commun.}, vol.~34,
  no.~2, pp. 318--328, Feb. 2016.

\end{thebibliography}
\end{NoHyper}
\end{document}